%% file: article.tex
\documentclass[a4paper,accepted=2022-10-07]{quantumarticle}

\pdfoutput=1
	
\input{preamble}

\begin{document}

\title{Simple and practical DIQKD security analysis via BB84-type uncertainty relations and Pauli correlation constraints}

\author{Michele Masini}
\email{michele.masini@ulb.be}

\author{Stefano Pironio}
\email{stefano.pironio@ulb.be}

\author{Erik Woodhead}
\email{erik.woodhead@ulb.be}

\affiliation{Laboratoire d'Information Quantique, Universit\'e libre de Bruxelles (ULB), Belgium}

\begin{abstract}
  According to the entropy accumulation theorem, proving the unconditional security of a device-independent quantum key distribution protocol reduces to deriving tradeoff functions, i.e., bounds on the single-round von Neumann entropy of the raw key as a function of Bell linear functionals, conditioned on an eavesdropper's quantum side information. In this work, we describe how the conditional entropy can be bounded in the 2-input/2-output setting, where the analysis can be reduced to qubit systems, by combining entropy bounds for variants of the well-known BB84 protocol with quantum constraints on qubit operators on the bipartite system shared by Alice and Bob. The approach gives analytic bounds on the entropy, or semi-analytic ones in reasonable computation time, which are typically close to optimal. We illustrate the approach on a variant of the device-independent CHSH QKD protocol where both bases are used to generate the key as well as on a more refined analysis of the original single-basis variant with respect to losses. We obtain in particular a detection efficiency threshold slightly below 80.26\%.
\end{abstract}

\maketitle

\section{Introduction\label{sec:twobasis}}

Based on Bell's theorem \cite{ref:b1964,ref:bc2014}, device-independent quantum key distribution (DIQKD) aims to allow cryptographic keys to be generated and proved secure based on minimal assumptions about the quantum devices \cite{ref:ab2007}. Following its proposal fifteen years ago, realizing a working DIQKD protocol has long presented a significant challenge both to theorists, due to the mathematical difficulty of devising practical and rigorous security proofs, and to experimental researchers, due to the difficulty of distributing entangled quantum systems with low noise and high detection rates over long distances. Recent advances paved the way to three successful proof-of-principle experiments demonstrating the feasibility of this technology  \cite{ref:nadlinger2022experimental,ref:zhang2022experimental,ref:liu2022toward}. However, there is still a long way from these proof-of-principle experiments to practical DIQKD implementations, with the necessity to improve the distance and the rate at which the keys are distributed.

One major theoretical advance introduced a few years ago is the entropy accumulation theorem \cite{ref:afd2018}, and the related technique of quantum probability estimation \cite{ref:zfk2020}, which reduces proving the unconditional security of a generic DIQKD protocol in the finite-key regime to the problem of obtaining a lower bound (called a \emph{min-tradeoff function} in \cite{ref:afd2018}) on the conditional von Neumann entropy $H(K_A | E)$ of Alice's raw key variable $K_A$ conditioned on an eavesdropper's possible quantum side information $E$, as a function of the expected value of a Bell expression. For instance the security of the simplest DIQKD protocol based on the CHSH inequality follows from the following lower bound on the conditional von Neumann entropy of Alice's measurement outcome $A_1$
\begin{equation}
  \label{eq:hae-chsh}
  H(A_1 | E)
  \geq 1 - \phi \bigro{\sqrt{S^{2}/4 - 1}} \,,
\end{equation}
where $\phi(x) = h\bigro{\tfrac{1}{2} + \tfrac{1}{2} x}$, $h(x)$ is the binary entropy, and
$S = \avg{A_{1} B_{1}} + \avg{A_{1} B_{2}} + \avg{A_{2} B_{1}} - \avg{A_{2}
  B_{2}}$ is the expected value of the CHSH Bell expression
\cite{ref:ab2007}.

The basic CHSH protocol based on the above lower bound is, however, not
optimal in a number of respects.  There has thus been in the last few years a
search for ways to bound the conditional entropy for more general DIQKD
protocols, either focusing on the 2-input/2-output setting
\cite{ref:tan2021computing,ref:sg2021,ref:ts2020}, or finding numerical methods to tackle the problem in a more general way \cite{ref:bff2021,ref:bff2021b}. Despite these efforts, bounding the entropy can be
a numerically-intensive problem, with one recent approach \cite{ref:ts2020}
notably requiring thousands of processor core-hours of computing time to
numerically bound the average entropy for a two-basis variant
\cite{ref:sg2021} of the CHSH-based DIQKD protocol. This has significant
drawbacks, reducing confidence in the results (as they are harder for others
to reproduce), increasing the difficulty to optimize over parameters in
simulations, and generally increasing the time and computing resources
necessary just to calculate a key rate. 

In this work, we present a new and versatile approach to bound the conditional entropy in the 2-input/2-output device-independent setting that is conceptually and technically relatively simple. It is a generalization of the approach in \cite{ref:wap2021} that was used to derive an analytical bound on the conditional entropy for a family of asymmetric CHSH inequalities. As we explain here, the main conceptual steps of this security analysis are not specific to the protocol considered in \cite{ref:wap2021} but can actually be easily adapted to other 2-input/2-output device-independent protocols.

The starting point is, as usual in the 2-input/2-output scenario, to use Jordan's lemma to reduce the analysis to convex combinations of qubit strategies. From there, our approach is based on three steps. First, as in a standard qubit QKD protocol like BB84, we bound the conditional entropy of Alice's key generating measurement, say, $A_1$ through an uncertainty relation involving the correlations $\avg{\bar{A_1} \otimes B}$ between an \emph{orthogonal} measurement $\bar{A_1}$ on Alice's subsystem and a binary observable $B$ on Bob's system. In a device-independent setting, though, and contrarily to, e.g., BB84, we cannot have direct access to the correlations $\avg{\bar{A_1}\otimes B}$ as we cannot assume that Alice's measurement devices perform measurements in two orthogonal bases $A_1$, $\bar{A_1}$. The second step is then to establish a device-independent qubit constraint on $\avg{\bar{A_1}\otimes B}$ which is based on correlations between Alice and Bob that are actually observed in the protocol, e.g., the CHSH expectation value or some other Bell expression. Combining the first and second step, we obtain a bound on the conditional entropy which is device-independent, apart from the assumptions that Alice and Bob are measuring qubits. The third step then involves a convexity analysis: either the resulting bound happens to be convex or, if this is not the case, we convexify it. In this way, we get a lower bound that is valid for convex combination of qubit strategies, and thus by Jordan's lemma, for arbitrary, dimension-free strategies.

We illustrate this new approach in detail on two variants of the CHSH-based DIQKD protocol: the two-basis variant \cite{ref:sg2021} and a new variant that incorporates, in addition to the CHSH value, information about the bias in the key generating measurement $A_1$. This last feature is particularly relevant for photonic implementations of DIQKD where no-click outcomes $\emptyset$ are mapped to a given key bit value, say $\emptyset \mapsto +1$, resulting in highly biased outcomes. The bounds that we obtain are optimal or close to optimal and significantly simpler technically and less computationally demanding than other approaches. We show in particular that a qubit DIQKD protocol can tolerate detector efficiencies as low as $80.26\%$.

We first provide in Section~\ref{sec:approach} a high-level description of our approach to bounding the conditional entropy in 2-input/2-output scenarios and then illustrate it in detail on the two-basis variant of the CHSH DIQKD protocol in Section~\ref{sec:two-basis} and on the variant optimized for losses in Section~\ref{sec:bias}.

\section{Description of our approach}
\label{sec:approach}

We start by specifying the class of problems that we aim to solve. 
We consider a tripartite setup involving a state $\rho_{ABE}$ shared among Alice, Bob, and the eavesdropper Eve. We assume that Alice can measure one of two $\pm 1$-valued observables $A_1$ or $A_2$ on her system, and similarly Bob can measure one of two $\pm 1$-valued observables $B_1$ or $B_2$. We refer to the tuple $\mathcal{Q}\equiv(\rho_{ABE},A_1,A_2,B_1,B_2)$ as a \emph{strategy}.

A strategy $\mathcal{Q}$ can be seen as describing a single round of a multi-round DIQKD protocol. The measurements by Alice and Bob serve two purposes: generating some random variable $K_A$ on Alice's side (which will constitute Alice's copy of the \emph{raw key} in the DIQKD protocol) and establishing some correlations between Alice and Bob (which will be estimated in a \emph{parameter estimation} step of the DIQKD protocol). Any strategy $\mathcal{Q}$ implies some tradeoff between how random $K_A$ is to Eve and how correlated Alice's and Bob's measurement outcomes are. This tradeoff can be formalized as follows.

\paragraph*{Eve's information on the raw key $K_A$.}
Let us assume that Alice uses the following general procedure to generate a random key value $K_A$: she first selects a measurement choice $X = 1,2$ according to a probability distribution $\mu_X$, she measures the corresponding observable $A_1$ or $A_2$, she gets the classical output $A = \pm 1$, and finally she applies to $A$ a (possibly stochastic) map $\$_x : \{\pm 1\} \to \mathcal{K}_A : A \mapsto K_A$ to obtain a value $K_A$ in some finite alphabet $\mathcal{K}_A$. A measure of how random $K_A$ is to Eve, given knowledge of the measurement choice $X$, is the conditional von Neumann entropy
\begin{equation}
  \label{eq:HKAXE}
  H(K_A | XE) = H(\rho_{K_AXE}) - H(\rho_{XE})  
\end{equation}
where $H(\rho) = -\Tr[\rho \log_{2}(\rho)]$ is the von Neumann entropy and $\rho_{XE} = \Tr_{K_A}[\rho_{K_AXE}]$ where
\begin{equation}
  \rho_{K_AXE} = \sum_{k_A,x} \mu(x) \proj{k_A,x} \otimes \rho^{k_{A},x}_{E}
\end{equation}
is the classical-quantum state describing the correlations between $K_A$, $X$, and $E$. In the above expression, the reduced states of Eve are given by
\begin{multline}
  \rho_{E}^{k_A,x} = \sum_{a = \pm 1} p_x(k_A | a)   \\
   \Tr_{AB} \biggsq{
    \rho_{ABE} \, \frac{\id + a A_x}{2} \otimes \id_B \otimes \id_E}
\end{multline}
where $p_x(k | a)$ are the transition probabilities of the map~$\$_x$. 

In this paper, we will often be interested in the case where $K_A$ is simply obtained as the outcome of one of Alice's  measurement, e.g., $A_1$ (i.e., there is no random input choice $X$ and no classical preprocessing.) By a slight abuse of notation, we write $A_1$ both for the random variable denoting the measurement outcome of $A_1$ and for the measurement $A_1$ itself. We thus write in such cases $K_A = A_1$ and $H(K_A | X E) = H(A_1 | E)$. We will also consider noisy preprocessing \cite{ref:rgk2005,ref:kr2008}, where Alice's raw key bit $K_A$ is again the outcome of the measurement $A_1$, but with probability $q$ she flips it and with probability $1 - q$ she keeps it as it is. We write $K_A = A_1^q$ for the corresponding random variable and thus $H(K_A | XE) = H(A^q_1 | E)$ for the conditional entropy. Finally, the last case we will consider is one where $K_A$ is obtained by choosing the observables $A_1$ and $A_2$ with probabilities $p$ and $\bar{p} = 1 - p$, respectively, and applying noisy preprocessing with flip probability $q$ to the measurement output. We then write $K_A=A_X^q$ and $H(K_A|XE)=H(A^q_X|XE)$.

\paragraph*{Alice-Bob correlations.}
In a device-independent setting, the correlations between Alice and Bob can be characterized through \emph{Bell linear functionals}, which are linear functions of 1-body and 2-body correlators. In the 2-input/2-output scenario, 1-body and 2-body correlators can all be written in the common form 
\begin{equation}\label{eq:corr}
  \avg{A_x \otimes B_y} = \Tr \bigsq{\rho_{AB} \, A_x \otimes B_y}
  \quad \text{for } x = 0,1,2
\end{equation}
if we define $A_0=\id_A$ and $B_0=\id_B$. A Bell linear functional $S$ is then specified by 9 real coefficients $\{S_{xy}\}_{x,y=0,1,2}$ $(x,y=0,1,2)$ and its value on a given set of correlators $\{\langle A_x\otimes B_y\rangle\}$ is given by
\begin{equation}
  S = \sum_{x,y=0}^2 S_{xy} \avg{A_x \otimes B_y} \,.
\end{equation}
We refer to $S$ as a \emph{Bell expectation}. We will particularly be interested in the following in the CHSH functional
\begin{equation}\label{eq:chsh}
  S = \avg{A_1 B_1} + \avg{A_1 B_2} + \avg{A_2 B_1} - \avg{A_2 B_2} \,.
\end{equation}

\paragraph*{Tradeoff between Eve's information on the raw key and Alice-Bob correlations.} 
Assume that a procedure for generating a raw key value (as specified by a measurement probability distribution $\mu_X$ and preprocessing maps $\$_x$) and a series of $m \geq 1$ Bell expectation values $\vect{S}=(S_1,\dotsc,S_m)$\footnote{This can range from a single Bell functional, such as CHSH, to the entire set of correlators $\{\avg{A_x \otimes B_y}\}$, or anything in between.} are fixed. Our objective is to establish a lower bound 
\begin{equation}\label{eq:hbound}
  H(K_A|XE) \geq f(\vect{S})
\end{equation}
that is device independent, in the sense that it is satisfied by every quantum strategy $\mathcal{Q}$. For technical reasons, we require $f$ to be a convex function of its arguments\footnote{This is required for application of the entropy accumulation theorem, and follows naturally when reducing the analysis to qubits. Furthermore, if $f$ defines a bound on $H(K|XE)$ that is tight, it must necessarily be convex by concavity of the conditional entropy and because any convex mixture of two strategies defines a valid strategy.}.

\paragraph*{Relation to the security of DIQKD protocols.}
In a typical DIQKD protocol, Alice's and Bob's devices are successively used for $n$ rounds. Some of the rounds are used to generate raw key values $K_A$ on Alice's side and $K_B$ on Bob's side. Some of the rounds are used to gather statistical data to decide, based on whether one or several Bell statistics are above some thresholds, if the protocol should be aborted or if it can proceed. In the latter case, error correction and privacy amplification are applied to the final raw key string. Following the application of the entropy accumulation theorem \cite{ref:afd2018}, the security of such a generic multi-round protocol can be reduced to deriving a tradeoff bound \eqref{eq:hbound}, which can be understood as characterizing the behavior of a single round\footnote{The raw key generation procedure and the set of Bell statistics to be used in the single-round bound \eqref{eq:hbound} should obviously coincide with those of the multi-round protocol.} in expectation. In particular a tradeoff bound allows one to compute the key rate in the finite-key regime and in the asymptotic one, where it simply reduces to the Devetak-Winter formula \cite{ref:dw2005}
\begin{equation}
  r = H(K_A | XE) - H(K_A | K_B) \,,  
\end{equation}
where $H(K_A | K_B)$ is the conditional Shannon entropy of the classical random variables $K_A$ and $K_B$.

\subsection{Reduction to qubits}

The lower bounds \eqref{eq:hbound} we aim to derive must be proven valid for any quantum strategy $\mathcal{Q}=(\rho_{ABE},A_1,A_2,B_1,B_2)$, defined a priori on Hilbert spaces of arbitrary dimension. However, because the strategies we consider involve only two binary measurements for Alice and for Bob, it is well-known that it is sufficient, thanks to Jordan's lemma, to consider pure qubit strategies \cite{ref:pa2009}.

More specifically, suppose that we have derived a lower bound $H(K_A|XE) \geq f(\vect{S})$, that is valid for any strategy $\mathcal{Q} = (\ket{\Psi}_{ABE},A_1,A_2,B_1,B_2)$ where \emph{i)} Alice's and Bob's systems are two-dimensional, \emph{ii)} $\ket{\Psi}_{ABE}$ is a pure state, \emph{iii)} $A_1$, $A_2,$ $B_1$, $B_2$ are qubit, non-degenerate Pauli observables constrained to the $\sz$--$\sx$ plane on the Bloch sphere, and where \emph{iv)} the function $f$ is convex. Then this lower bound is valid for arbitrary strategies. For details, see for instance \cite{ref:wap2021}.

Note that the ``2-input/2-output'' restriction, which allows to make this qubit simplification, only applies to Alice's measurements and to those measurements of Bob that are involved in the definition of the Bell functionals $\vect{S}$, as these are the only measurements involved in the relation \eqref{eq:hbound}. The raw key generation procedure on Bob's side leading to the raw key value $K_B$ can, however, involve further measurement choices with more outputs, see examples in the Section~\ref{sec:applications}.

We now assume the above simplification and present our approach to deriving tradeoff bounds, which follows three technical steps described in the next three subsections.

\subsection{BB84-type uncertainty relations}
\label{sec:bb84}

The first non-trivial step in our approach is \emph{device-dependent} and consists in deriving a qubit uncertainty relation akin to those used in the analysis of the standard entanglement-based BB84 protocol and variants of it. 
Let us illustrate this on several examples. In the following, $\phi(x) = h\bigro{\tfrac{1}{2} + \tfrac{1}{2} x}$, where $h(x)$ is the binary entropy.

Consider first the simple situation where Alice's raw key bit $K_A=A_1$ is simply obtained as the outcome of the measurement $A_1$, i.e., there is no random input choice $X$ and no classical preprocessing. We then have the following bound.
\begin{namedthm}{Entropy bound 1 (BB84)}
  \begin{equation}
    \label{eq:bb84-bound}
    H(A_1| E) \geq 1 - \phi \bigro{\abs{\avg{\bar{A}_1 \otimes B}}} \,,
  \end{equation}
  where $\bar{A}_1$ is a Pauli observable orthogonal to $A_1$ on the Bloch sphere and $B$ any given $\pm 1$-valued observable on Bob's subsystem.
\end{namedthm}
This bound is simply a reexpression of the one-sided device-independent entropy bound $H(\sz | E) \geq 1 - \phi \bigro{\abs{\avg{\sx \otimes B}}}$ for the BB84 protocol \cite{ref:bc2010} that relates the information Eve has about the outcome of a $\sz$ measurement by how much Bob is correlated to the complementary $\sx$ measurement. The bound \eqref{eq:bb84-bound} directly follows from the fact that $A_1$ and $\bar{A}_1$ are Pauli operators, which we can identify with the $\sz$ and $\sx$ operators. 

As a second example, let us add noisy preprocessing \cite{ref:rgk2005,ref:kr2008} to the raw key procedure: Alice's raw key bit $K_A=A_1^q$ is again the outcome of the measurement $A_1$, but with probability $q$ she flips it and with probability $1 - q$ she keeps it as it is.
\begin{namedthm}{Entropy bound 2 (BB84 bound with noisy preprocessing)}
  \begin{IEEEeqnarray}{rCl}
    \label{eq:HZE-bb84-noise}
    H(A^q_1| E) &\geq& f_q(\abs{\avg{\bar{A}_1 \otimes B}})\,,
  \end{IEEEeqnarray}
  where
  \begin{IEEEeqnarray}{rCl}\label{eq:fqx}
    f_q(x) & =& 1 + \phi \Bigro{\sqrt{(1 - 2 q)^{2}
        + 4 q (1 - q) x^{2}}} \nonumber \\
    &&-\> \phi(x) \,,
  \end{IEEEeqnarray}
  and $\bar{A}_1$ is a Pauli observable orthogonal to $A_1$ on the Bloch sphere and $B$ any given $\pm 1$-valued observable on Bob's subsystem.
\end{namedthm}  
This again follows by identifying $A_1$ and $\bar{A}_1$ with the $\sz$ and $\sx$ operators and reusing a one-sided device-independent bound known for BB84 with noisy preprocessing \cite{ref:w2014,ref:wap2021}.

The two above bounds were used in \cite{ref:wap2021} to analyze the security of a family of CHSH-based DIQKD protocols. But more generally, it is also possible to obtain other bounds, such as the two ones below, which we will apply to other variants of CHSH-based DIQKD protocols in Section~\ref{sec:applications}.

\begin{namedthm}{Entropy bound 3 (BB84 with noisy preprocessing and bias)}
  \begin{equation}
    \label{eq:HZE-A1-XB}
    H(A^q_1| E) \geq g_{q} \bigro{\abs{\avg{A_{1}}},
      \abs{\avg{\bar{A}_1 \otimes B}}} \,,
  \end{equation}
  where
  \begin{IEEEeqnarray}{rCl}
    g_{q}(z,x) &=& \phi \bigro{\tfrac{1}{2} (R_{+} + R_{-})}
    + \phi \bigro{\tfrac{1}{2} (R_{+} - R_{-})} \nonumber \\
    &&-\> \phi \bigro{\sqrt{z^{2} + x^{2}}} \,,
  \end{IEEEeqnarray}
  with
  \begin{equation}
    R_{\pm} = \sqrt{(1 - 2 q \pm z)^{2} + 4 q (1 - q) x^{2}} \,,
  \end{equation}
  and $\bar{A_1}$ is a Pauli observable orthogonal to $A_1$ on the Bloch sphere and $B$ any given $\pm 1$-valued observable on Bob's subsystem.
\end{namedthm}
This bound represents a refinement of the bound 2, as it depends not only on $\avg{\bar{A}_1 \otimes B}$, but also on the value of the 1-body correlator $\avg{A_{1}}$ measuring how much Alice's raw output is biased. 

Our last example is one where Alice's raw key bit $K_A = A_X^q$ is obtained by choosing the observables $A_1$ and $A_2$ with probability $p$ and $\bar{p} = 1 - p$, respectively, and applying noisy preprocessing with flip probability $q$ to the measurement output. The conditional entropy is then
\begin{equation}
  \label{eq:entropy-two-basis}
  H(A_X^q|XE)=pH(A_1^q|E)+\bar p H(A_2^q|E)\,,
\end{equation}
and one has the following bound.
\begin{namedthm}{Entropy bound 4 (Two-basis bound)}
  \begin{equation}
    \label{eq:twobasis-bound}
    H(A_X^q|XE) \geq f_q \Bigro{\sqrt{
        p \avg{\bar{A}_{1} \otimes B}^{2}
        + \bar{p} \avg{\bar{A}_{2} \otimes B'}^{2}}}
  \end{equation}
  where $\bar{A}_1$ and $\bar{A}_2$ are observables orthogonal to $A_1, A_2$, respectively and $f_q(x)$ is the function defined in \eqref{eq:fqx}.
\end{namedthm}

The above bounds are essentially similar to those used in the analysis of standard entanglement-based QKD. They are valid for arbitrary entangled states $\ket{\Psi}_{ABE}$ where Alice's and Bob's systems are two dimensional and are expressed in terms of correlators $\avg{A \otimes B}$ between Alice and Bob that involve (contrarily to the device-independent case) \emph{specific, fixed} observables, such as $\bar{A}_1$ on Alice's side. As such they can be derived using existing techniques.

We remark that all of these bounds can be derived from bound~3, which we derive in detail in Appendix~\ref{ap:part-sym}. In particular, bound~2 is a special a case of bound~3 evaluated with $\avg{A_{1}} = 0$\footnote{The resulting bound holds independently of the actual value of $\avg{A_{1}}$ thanks to the monotonicity property discussed below: if we make in bound~3 the replacement $\abs{\avg{A_1}} \mapsto 0$ we obtain a bound that remains valid.}, while bound~1 is obtained by further setting $q = 0$. Bound~4 follows from bounding both contributions to the average entropy separately using bound~2,
\begin{IEEEeqnarray}{rCl}
  H(A^q_X|XE) &=& p H_q(A_{1} | E) + \bar{p} H_q(A_{2} | E)  \\
  &\geq& p f_{q} \bigro{\abs{\avg{\bar{A}_{1} \otimes B}}}
  + \bar{p} f_{q} \bigro{\abs{\avg{\bar{A}_{2} \otimes B'}}} \nonumber \,,
\end{IEEEeqnarray}
and then using that the function $x \mapsto f_{q}(\sqrt{x})$ is convex (see Appendix~B of \cite{ref:wap2021} for a proof of this property).

Importantly, we also show in Appendix~\ref{ap:part-sym} that all the above bounds satisfy a type of monotonicity property.
We say that a bound $H(K_A|XE)\geq f(x)$ is \emph{monotone} in $x$ if the bound $H(K_A|XE)\geq f(x_{-})$ holds for all $x_{-}\leq x$ and similarly in the multivariate case for each variable independently, e.g., $H(K_A|XE)\geq f(x,y)$ is \emph{monotone} in $x$ and $y$ if the bound $H(K_A|XE)\geq f(x_{-},y_{-})$ hold for all $x_{-}\leq x$ and $y_{-}\leq y$. Note that this monotonicity property is weaker than monotonicity of the function $f$ itself: if the function $f$ is monotonically increasing then the bound $H(K_A|XE)\geq f(x)$ is monotone, but the converse does not necessarily hold.
\begin{namedthm}{Monotonicity property}
  The entropy bounds \eqref{eq:bb84-bound} and \eqref{eq:HZE-bb84-noise} are monotone in $\abs{\avg{\bar{A_1} \otimes B}}$, the bound \eqref{eq:HZE-A1-XB} is monotone in $\abs{\avg{A}_1}$ and $\abs{\avg{\bar{A}_1 \otimes B}}$, and the bound \eqref{eq:twobasis-bound} is monotone in $p \avg{\bar{A}_{1} \otimes B}^{2} + \bar{p} \avg{\bar{A}_{2} \otimes B'}^{2}$.  
\end{namedthm}
The monotonicity of the bound \eqref{eq:HZE-A1-XB} is established in Appendix~\ref{ap:part-sym} from which the monotonicity of the other bounds follows\footnote{In the case of bounds \eqref{eq:bb84-bound}, \eqref{eq:HZE-bb84-noise}, \eqref{eq:twobasis-bound}, it also follows from the stronger property that the function $f_q(x)$ is monotonically increasing in $x$, as shown in Appendix~B. of \cite{ref:wap2021}.}. 
This property will be important in Section~\ref{sec:correlations} as it allows replacing in the entropy bounds the correlators on which they depend in the right-hand side by a lower bound on these correlators and in Section~\ref{sec:convexity} where it allows the systematic computation of a convex envelope based on a discrete set of points.

\subsection{Pauli correlation constraints}
\label{sec:correlations}

The bounds on the conditional entropy $H(K_A|XE)$ that we have given in the previous subsection are expressed in terms of correlators involving observables which are not necessarily accessible through the devices, e.g., the correlator $\avg{\bar A_1 \otimes B}$ involving the observable $\bar A_1$. The second step of our approach consists in deriving a constraint on these correlators in terms of correlators involving only the observables $A_{1}$, $A_{2}$, $B_{1}$, $B_{2}$ \emph{actually measured} by the devices.

For instance, it is a straightforward exercise, see \cite{ref:wap2021}, to show the following bound.
\begin{namedthm}{Correlation bound 1 (CHSH)}
  \begin{equation}
    \label{eq:S-XB-bound}
    \abs{\avg{\bar A_1 \otimes B}} \geq \sqrt{S^2/4-1} \,,
  \end{equation}
  where $S = \avg{A_{1} B_{1}} + \avg{A_{1} B_{2}} + \avg{A_{2} B_{1}} - \avg{A_{2} B_{2}}$ is the expected value of the CHSH statistic and $B \propto B_{1} - B_{2}$.
\end{namedthm}

More generally, one can also consider a family of asymmetric versions of the CHSH statistic for which the following bounds are shown in \cite{ref:wap2021}.
\begin{namedthm}{Correlation bound 2 (asymmetric CHSH)}
  Let $S_\alpha = \alpha \avg{A_{1} B_{1}} + \alpha \avg{A_{1} B_{2}} + \avg{A_{2} B_{1}} - \avg{A_{2} B_{2}}$ be a variant of CHSH depending on a given parameter $\alpha\in\mathbb{R}$. Then for some appropriate choice of a $\pm 1$-valued observable $B$,
  \begin{equation}
    \label{eq:Salpha-XB-bound}
    \abs{\avg{\bar A_1\otimes B}} \geq E_\alpha(S_\alpha)\,,
  \end{equation}
  where 
  \begin{equation}
    E_\alpha(S_\alpha) = \sqrt{{S_\alpha}^2/4-\alpha^2}
  \end{equation}
  if $\abs{\alpha}\geq 1$ or $\abs{S_\alpha} \geq 2\sqrt{1+\alpha^2-\alpha^4}$ and
  \begin{equation}
    E_\alpha(S_\alpha) = \sqrt{1 - \Bigro{1 - \tfrac{1}{\abs{\alpha}} \sqrt{(1 - \alpha^{2}) ({S_{\alpha}}^{2}/4 - 1)}}^{2}}
  \end{equation}
  otherwise.
\end{namedthm}

The correlation bounds \eqref{eq:S-XB-bound} and \eqref{eq:Salpha-XB-bound}  can be derived analytically. But more generically, one can derive numerical lower bounds on polynomial functions of arbitrary qubit correlators, such as $\avg{\bar A_1 \otimes B}$ or $\avg{\bar A_2 \otimes B'}$, in terms of Bell functionals involving only the accessible correlators $\avg{A_x \otimes B_y}$ ($x,y=0,1,2$), using the Lasserre hierarchy of semidefinite programming relaxations for polynomial optimization \cite{ref:l2001,henrion2006}. This can be done by parameterizing explicitly all qubit operators in the $\sz$--$\sx$ plane.

We illustrate this general idea on the specific problem of deriving a lower bound for the expression 
\begin{equation}
  \label{eq:twobasis-corr}
  p \avg{\bar{A}_{1} \otimes B}^{2}
  + \bar{p} \avg{\bar{A}_{2} \otimes B'}^2
\end{equation}
appearing on the right-hand side of \eqref{eq:twobasis-bound} in terms of the CHSH expectation value $S$.

We first recall that we can use any $\pm 1$-valued observables $B$ and $B'$ in \eqref{eq:twobasis-bound}. Taking these to be of the form
\begin{equation}
  B^{(\prime)} = \cos \bigro{\varphi^{(\prime)}_{B}} \sz
  + \sin \bigro{\varphi^{(\prime)}_{B}} \sx
\end{equation}
and then choosing the angles $\varphi_{B}$ and $\varphi'_{B}$ that maximize  \eqref{eq:twobasis-corr} we obtain
\begin{IEEEeqnarray}{rCl}
  \IEEEeqnarraymulticol{3}{l}{
    p \avg{\bar{A}_{1} \otimes B}^{2}
    + \bar{p} \avg{\bar{A}_{2} \otimes B'}^2} \nonumber \\
  \qquad &=& p \bigro{
    \avg{\bar{A}_{1} \otimes \sz}^{2}
    + \avg{\bar{A}_{1} \otimes \sx}^{2}} \nonumber \\
  &&+\> \bar{p} \bigro{
    \avg{\bar{A}_{2} \otimes \sz}^{2}
    + \avg{\bar{A}_{2} \otimes \sx}^{2}} \,.
\end{IEEEeqnarray}
We then choose Alice's basis such that
\begin{IEEEeqnarray}{rCl}
  A_{1} &=& \cos \bigro{\tfrac{\varphi_A}{2}} \sz
  - \sin \bigro{\tfrac{\varphi_A}{2}} \sx \,, \\
  A_{2} &=& \cos \bigro{\tfrac{\varphi_A}{2}} \sz
  + \sin \bigro{\tfrac{\varphi_A}{2}} \sx
\end{IEEEeqnarray}
and the complementary operators are
\begin{IEEEeqnarray}{rCl}
  \bar{A}_{1} &=& \sin \bigro{\tfrac{\varphi_A}{2}} \sz
  + \cos \bigro{\tfrac{\varphi_A}{2}} \sx \,, \\
  \bar{A}_{2} &=& - \sin \bigro{\tfrac{\varphi_A}{2}} \sz
  + \cos \bigro{\tfrac{\varphi_A}{2}} \sx
\end{IEEEeqnarray}
for some unknown angle $\varphi_{A}$. Using these in the above expression we obtain, explicitly,
\begin{IEEEeqnarray}{rCl}
  \IEEEeqnarraymulticol{3}{l}{
    \label{eq:objective-in-paulis}
    p \avg{\bar{A}_{1} \otimes B}^{2}
    + \bar{p} \avg{\bar{A}_{2} \otimes B'}^{2}} \nonumber \\
   \ &=& \sin \bigro{\tfrac{\varphi_A}{2}}^{2}
  (E\du{\rzz}{2} + E\du{\rzx}{2})
  + \cos \bigro{\tfrac{\varphi_A}{2}}^{2}
  (E\du{\rxz}{2} + E\du{\rxx}{2}) \nonumber \\
  &&+\> 2(2p-1) \sin \bigro{\tfrac{\varphi_A}{2}}
  \cos \bigro{\tfrac{\varphi_A}{2}} (E_{\rzz} E_{\rxz} + E_{\rzx} E_{\rxx}) \,,
  \nonumber \\*
\end{IEEEeqnarray}
where we note the expectation values of products of Pauli operators $E_{\rxx} = \avg{\sx \otimes \sx}$ and similarly for $E_{\rxz}$, $E_{\rzx}$, and $E_{\rzz}$.

We wish to constrain \eqref{eq:objective-in-paulis} for a given value of the CHSH expectation value which, in the choice of basis made above, takes the form
\begin{IEEEeqnarray}{rCl}
  S &=& \avg{(A_1 + A_2) \otimes B_1} + \avg{(A_1 - A_2) \otimes B_2}
  \nonumber \\
  &=& 2 \cos \bigro{\tfrac{\varphi_{A}}{2}} \avg{\sz \otimes B_{1}}
  - 2 \sin \bigro{\tfrac{\varphi_{A}}{2}} \avg{\sx \otimes B_{2}} \,.
  \IEEEeqnarraynumspace
\end{IEEEeqnarray}
Maximizing the second line over (nondegenerate) $\pm 1$-valued observables $B_{1}$ and $B_{2}$ in the $\sz$--$\sx$ plane gives
\begin{IEEEeqnarray}{rCl}
  \label{eq:chsh-pauli-constraint}
  S/2 &\leq& \abs{\cos \bigro{\tfrac{\varphi_A}{2}}}
  \sqrt{E\du{\rzz}{2} + E\du{\rzx}{2}} \nonumber \\
  &&+\> \abs{\sin \bigro{\tfrac{\varphi_A}{2}}}
  \sqrt{E\du{\rxz}{2} + E\du{\rxx}{2}} \,,
\end{IEEEeqnarray}
which can be read as a constraint on the unknown angle $\varphi_A$ and Pauli correlations $E_{\rxx}$, $E_{\rxz}$, $E_{\rzx}$, and $E_{\rzz}$ appearing in \eqref{eq:objective-in-paulis}.

To complete the problem, we finally remark that $E_{\rxx}$, $E_{\rxz}$, $E_{\rzx}$, and $E_{\rzz}$ can be interpreted as expectations of products of the $\sz$ and $\sx$ Pauli operators for some underlying state only if they satisfy
\begin{IEEEeqnarray}{rCl}
  E\du{\rzz}{2} + E\du{\rzx}{2} &\leq& 1 \,, \label{eq:pauli-constraint-1}\\
  E\du{\rxz}{2} + E\du{\rxx}{2} &\leq& 1 \,,\label{eq:pauli-constraint-2}
\end{IEEEeqnarray}
and
\begin{IEEEeqnarray}{l}
  \label{eq:pauli-constraint-3}
  \bigro{1 - E\du{\rzz}{2} - E\du{\rzx}{2}}
  \bigro{1 - E\du{\rxz}{2} - E\du{\rxx}{2}} \nonumber \\
  \qquad \geq\> \bigro{E_{\rzz} E_{\rxz} + E_{\rzx} E_{\rxx}}^{2}
\end{IEEEeqnarray}
as shown in Section~4.3 of \cite{ref:wap2021}.
	
To get a valid lower bound on \eqref{eq:two-basis-corr}, it is thus sufficient to minimize the left-hand side of \eqref{eq:objective-in-paulis} given the constraints \eqref{eq:chsh-pauli-constraint}--\eqref{eq:pauli-constraint-3}. The problem can be simplified by introducing the new variables
\begin{IEEEeqnarray}{rCl+rCl}
  E_{\rzz} &=& \lambda \cos(z) \,, & E_{\rzx} &=& \lambda \sin(z) \,, \\
  E_{\rxz} &=& \mu \cos(x) \,, & E_{\rxx} &=& \mu \sin(x) \,, \\
  s &=& \sin \bigro{\tfrac{\varphi_A}{2}} \,, &
  c &=& \cos \bigro{\tfrac{\varphi_A}{2}} \,, \IEEEeqnarraynumspace \\
  \Delta &=& \cos(x-z) \,. &&&
\end{IEEEeqnarray}
Using the trigonometric identity $\cos \bigro{\tfrac{\varphi_A}{2}}^{2} + \sin  \bigro{\tfrac{\varphi_A}{2}}^{2} = 1$ and that we can drop the absolute values from \eqref{eq:chsh-pauli-constraint} without substantially changing the problem, we arrive at the following.
\begin{namedthm}{Correlation bound 3 (two-basis)}
  There exist $\pm 1$-valued qubit operators $B$ and $B'$ acting on Bob's subsystem such that
  \begin{equation}
    \label{eq:two-basis-corr}
    p \avg{\bar{A}_{1} \otimes B}^{2}
    + \bar{p} \avg{\bar{A}_{2} \otimes B'}^{2} \geq E_p(S)^{2} \,,
  \end{equation}
  where $E_p(S)^{2}$ is the solution to the minimization problem
  {\normalfont
    \begin{IEEEeqnarray*}{Ru+rCl}
      \label{eq:twobasis-polyopt}
      E_p(S)^{2} = & min & \IEEEeqnarraymulticol{3}{l}{
        s^2 \lambda^2 +c^2 \mu^2 + 2(2p-1) s c \lambda \mu \Delta}
      \nonumber \\
      & s.t. & c \lambda + s \mu &\geq& S/2 \\
      &&\lambda^2 &\leq& 1 \\
      &&\mu^2 &\leq& 1 \\
      &&(1 - \lambda^2) (1 - \mu^2) &\geq& \lambda^2 \mu^2 \Delta^2 \\
      &&c^2+s^2 &=& 1 \\
      &&\Delta^2 &\leq& 1 \IEEEyesnumber
    \end{IEEEeqnarray*}}
  in the five variables $\lambda, \mu, c, s, \Delta \in \mathbb{R}$.
\end{namedthm}
As the above is a polynomial optimization problem, it can be reduced to a sequence of semidefinite programs using the Lasserre hierarchy \cite{ref:l2001,henrion2006}. Importantly, every SDP relaxation at a given order in the hierarchy provides a valid lower bound to the optimization problem and consequently a valid lower bound of the form \eqref{eq:two-basis-corr}. At level 3 of the Lasserre hierarchy, the problem takes less than a second to solve and appears to already give the optimal solution. 

In the case in which $p = 1/2$, the above problem can actually be solved analytically, as shown in Appendix~\ref{ap:two-basis}. The result in that case is
\begin{equation}
  \label{eq:equiv}
  E_{\frac{1}{2}}(S)^{2} = \frac{1 + x\du{*}{2}}{1 - x_{*}}
  + \frac{S^2}{4} \frac{1 + x_{*}}{1 - x_{*}}
  - \frac{S}{\sqrt{2}} \frac{(1 + x_{*})^{3/2}}{1 - x_{*}} \,,
\end{equation}
where the variable $x_{*}$ is the solution of
\begin{equation}
  \label{eq:to-solve}
  4 x (2 - x) + 2(S^2 + 2) + S (x - 5) \sqrt{2 (1 + x)} = 0
\end{equation}
in the range	
\begin{equation}
  \label{eq:an-bounds}
  -\frac{S}{4} \sqrt{8 - S^2} \leq x \leq \frac{S}{4} \sqrt{8 - S^2} \,.
\end{equation} 
Eq.~\eqref{eq:to-solve} can be rearranged to a root-finding problem for a degree 4 polynomial in $x$ and can thus be solved analytically, though the solution is quite lengthy and we do not explicitly report it here.

\subsection{Convexity and fully device-independent bounds}
\label{sec:convexity}

Combining the above correlation bounds and the entropy bounds of the previous section, one obtains bounds on the conditional entropy that are device independent modulo the qubit reduction. For instance, using the CHSH correlation bound \eqref{eq:S-XB-bound} in the BB84 entropy bound \eqref{eq:bb84-bound}, where the substitution of \eqref{eq:S-XB-bound} in \eqref{eq:bb84-bound} is possible thanks to the monotonicity property of the BB84 entropy bound, we recover the CHSH entropy bound
\begin{equation}
  \label{eq:chsh-bound}
  H(A_1 | E) \geq 1 - \phi \bigro{\sqrt{S^2/4 - 1}}
\end{equation}
given in the introduction and originally derived in \cite{ref:ab2007}. Using \eqref{eq:Salpha-XB-bound} in the BB84 bound with noisy preprocessing \eqref{eq:HZE-bb84-noise}, one obtains the more general qubit bound
\begin{equation}
  \label{eq:salpha-qubit-bound}
  H(A^q_1 | E) \geq f_{q} \bigro{E_\alpha(S_\alpha)}
\end{equation}
derived in \cite{ref:wap2021}.

But other combinations are also possible, such as the two original following ones, which we are going to consider in more detail in Section~\ref{sec:applications}. 

The first, which gives a bound on the entropy in terms of $\avg{A_{1}}$ in addition to CHSH, is simply obtained by combining \eqref{eq:S-XB-bound} and \eqref{eq:HZE-A1-XB}:
\begin{equation}
  \label{eq:bias}
  H(A^q_1 | E) \geq g_{q} \bigro{\abs{\avg{A_{1}}}, \sqrt{S^2/4 - 1}} \,.
\end{equation}

For the second, let $\tilde E_p(S)^{2}$ denote any lower bound to $E_p(S)^2$ obtained by solving analytically or numerically the polynomial optimization problem \eqref{eq:twobasis-polyopt} or any of its relaxations in the Lasserre hierarchy. Then using such a bound in \eqref{eq:twobasis-bound}, we obtain
\begin{equation}
  \label{eq:HAXE-twobasis-qubit}
  H(A^q_X | XE) \geq f_{q} \bigro{\tilde E_p(S)}
\end{equation}
with $\tilde E_{p}(S) \equiv \sqrt{\tilde E_{p}(S)^{2}}$.

\subsubsection{Convexity analysis}

Regardless of the combination used, the result is a bound on the conditional entropy valid for two-qubit systems, which can only be extended to give a fully device-independent bound, valid in arbitrary dimension, if it is convex. The third and final step thus consists of a convexity analysis.

If we obtain a qubit bound on the conditional entropy with a reasonably simple analytic expression then it may be feasible to study its properties directly. Either we simply prove it is convex, as can be done for \eqref{eq:chsh-bound}, or more generally as was done in \cite{ref:wap2021} for \eqref{eq:salpha-qubit-bound} for $\abs{\alpha}\geq 1$. Or we analytically establish that it is not convex and determine its convex envelope, as was done in \cite{ref:wap2021} for \eqref{eq:salpha-qubit-bound} for $\abs{\alpha} < 1$.

More generally, however, the qubit bound may be obtained numerically or it may be analytic but of a form that does not easily lend itself to an analytic convexity analysis, as is the case for the bounds \eqref{eq:bias} and \eqref{eq:HAXE-twobasis-qubit}. In such cases, we need a way of constructing a convex lower bound on whatever qubit bound we obtain.

\subsubsection{Convex lower bounds through linear programming}
\label{sec:linear-programming}

A simple solution that we can use, provided our entropy bounds satisfy the monotonicity property introduced in subsection~\ref{sec:bb84}, is based on a discretization of the qubit bound, similar to the approach used in \cite{ref:sg2021}. In the following, let us generically write the bound valid for two-qubit systems as
\begin{equation}
  \label{eq:two-qubit}
  H(K_{A} | X E) \geq f(\vect{S}) \,,
\end{equation}
where $f \colon \mathcal{D} \to \mathbb{R}$ is a function, defined on some domain $\mathcal{D}$, that we either know analytically or can compute numerically, of one or more Bell expectation values $\vect{S} = (S_{1}, S_{2}, \dotsc, S_{n}) \in \mathcal{D}$.

Let us introduce a covering $\mathcal{K} = \{K\}$ of the domain $\mathcal{D}$ by polytopes $\{K\}$, such that every $\vect{S} \in \mathcal{D}$ is contained in at least one of the polytopes $K$. In practice, we would typically use a grid partition in terms of hyperrectangles where each point (outside of vertices and shared edges) is contained in only one hyperrectangle $K$ (but this is not strictly necessary for the method to work).

Let us suppose, furthermore, that for every $K$ we have a way of identifying a value $f[K]$ that we can use as a lower qubit bound on the conditional entropy valid for the entire polytope, i.e., such that
\begin{IEEEeqnarray}{c+c}
  \label{eq:f[k]}
  H(K_{A} | X E) \geq f[K] \,, &
  \forall \vect{S} \in K \,.
\end{IEEEeqnarray}
We can then define a discretized qubit bound,
\begin{equation}
    H(K_{A} | X E) \geq f_{\mathcal{K}}(\vect{S})
\end{equation}
where $f_{\mathcal{K}}$ is defined as
\begin{equation}
  \label{eq:discretized-f-def}
  f_{\mathcal{K}}(\vect{S}) = \min_{K \ni \vect{S}} f[K] \,,
\end{equation}
where the minimization is taken over all polytopes $K$ that contain $\vect{S}$. This, in particular, associates unique values $f_{\mathcal{K}}(\vect{S}_{j})$ to the vertices $\vect{S}_{j}$ of the polytopes. The convex envelope of the discretized function $f_{\mathcal{K}}$, finally, is readily given by the solution to the following linear programming problem,
\begin{IEEEeqnarray}{curCl}
  \bar{f}_{\mathcal{K}}(\vect{S}) =\>
  &minimize & \IEEEeqnarraymulticol{3}{l}{
    \sum_{j} \theta_{j} \,
    f_{\mathcal{K}}(\vect{S}_{j}}) \nonumber \\
  &subject to &
  \sum_{\vect{j}} \theta_{j} \vect{S}_{j} &=& \vect{S} \nonumber \\
  && \sum_{j} \theta_{j} &=& 1 \nonumber \\
  && \theta_{j} &\geq& 0 \,,
\end{IEEEeqnarray}
where the $\vect{S}_{j}$ are the combined vertices of all the polytopes $K$ in $\mathcal{K}$.
 We thus obtain a bound
\begin{equation}
  H(K_{A} | X E) \geq \bar{f}_{\mathcal{K}}(\vect{S})
\end{equation}
on the conditional entropy that is convex and extends to the fully device-independent setting.

We have not explained, however, how one can identify in \eqref{eq:f[k]} the lower-bound values $f[K]$ for each polytope $K$, which is crucial to define a discretized qubit bound. This can be done if the bound \eqref{eq:two-qubit} is monotone in $\abs{\vect{S}} = (\abs{S_{1}}, \abs{S_{2}}, \dotsc, \abs{S_{n}})$, i.e., if the bound still holds if we replace in \eqref{eq:two-qubit} any of the $n$ Bell expectation values $S_i$ by a value $s_i$ that is smaller in absolute value, $\abs{s_i}\leq \abs{S_i}$. This is in particular the case for all the bounds \eqref{eq:chsh-bound}--\eqref{eq:HAXE-twobasis-qubit} presented above since they are obtained by combining the monotone entropy bounds of subsection~\ref{sec:bb84} with the monotonically increasing correlation bounds of subsection~\ref{sec:correlations}. Using this monotonicity property, we can now simply divide the domain $\mathcal{D}$ into hyperrectangles $K$ and use as the lower-bound value $f[K]$ for each hyperrectangle $K$, the value of the qubit bound evaluated  at the corner that is closest to the origin.

\begin{figure}[tbp]
  \centering
  \includegraphics{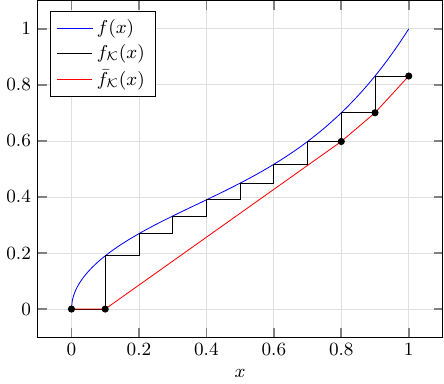}
  \caption{Convex lower bound $\bar{f}_{\mathcal{K}}$ of a function $f$ constructed on $n$ (in the figure $n=10$) equally spaced subdivisions of its domain, i.e., the polytopes $K$ are here $n$ consecutive line segments between $x=0$ and $x=1$. We actually used this method on the qubit bound \eqref{eq:HAXE-twobasis-qubit}, but the function $f_q(\tilde{E}_p(S))$ is too close to convex to make a visually interesting example. The construction is thus illustrated on the figure for the visibly non-convex function $f(x) = 0.6 \sqrt{x} + 0.4 x^{4}$. }
  \label{fig:chull}
\end{figure}

Finally, in the special case that we are working with a qubit entropy bound $H(K_{A} | X E) \geq f(S)$ of a single variable $S$, we remark that one can avoid the linear program and compute $f_{\mathcal{K}}(S)$ very rapidly essentially by eliminating the redundant vertices and interpolating between the remaining ones, as illustrated in Figure~\ref{fig:chull}. This can be done in linear time in the number of vertices \cite{ref:mca1979,ref:svw1987}. We in particular applied this technique to the two-basis bound \eqref{eq:HAXE-twobasis-qubit} to compute the key-rate bounds obtained in Section~\ref{sec:two-basis} below.

\subsubsection{Certifying an affine tradeoff bound}
\label{sec:affine-tradeoff}

While we can always use the above approach when we have a qubit entropy bound satisfying the monotonicity property, it is not always necessary to solve the linear programming problem to obtain a valid convex lower bound on the conditional entropy. An alternative approach, which would ultimately lend itself to more direct use in the entropy accumulation theorem, is to certify a linear or affine lower bound on the entropy.

Here, let us suppose we believe that the conditional entropy respects an affine lower bound
\begin{equation}
  \label{eq:tangent}
  H(K_{A} | X E) \geq \beta + \vect{\alpha} \cdot \vect{S} - \varepsilon \,,
\end{equation}
that we wish to certify up to some precision $\varepsilon$. Such a bound may be obtained, for example, by computing at a particular point the tangent of a function $\bar{f}(\vect{S})$ that we believe to be the convex hull of a known qubit bound $f(\vect{S})$. As above, we introduce a covering $\mathcal{K} = \{K\}$ of the domain $\mathcal{D}$ with polytopes $K$ and assume for every $K$ a lower bound $f[K]$ on the conditional entropy, as defined in \eqref{eq:f[k]}.  We also define
\begin{IEEEeqnarray}{rCl}
  \alpha[K]
  &=& \max_{\vect{S} \in K} \, \vect{\alpha} \cdot \vect{S} \nonumber \\
  &=& \max_{\vect{S} \in \verx(K)} \, \vect{\alpha} \cdot \vect{S}
\end{IEEEeqnarray}
where $\verx(K)$ are the vertices of $K$. To check that \eqref{eq:tangent} holds, we then only need to verify that
\begin{equation}
  \label{eq:tangent-test}
  \beta + \alpha[K] - f[K] \leq \varepsilon
\end{equation}
holds for all polytopes $K$ in the covering $\mathcal{K}$, which is now a finite problem. Alternatively, we can compute the maximal value over $\mathcal{K}$ of $\beta + \alpha[K] - f[K]$ to determine the best possible precision $\varepsilon$ we can achieve given our covering choice. 

\begin{figure}[t]
  \includegraphics[scale=0.98]{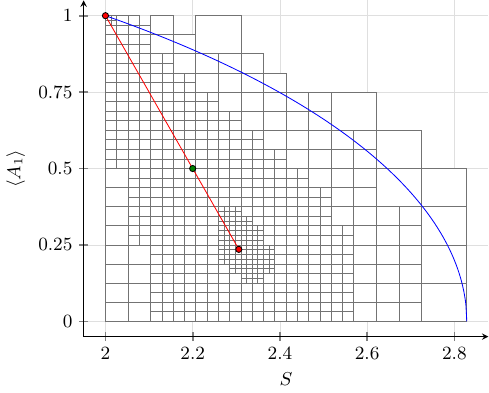}
  \caption{Certification of an affine lower entropy bound based on the qubit bound \eqref{eq:bias} depending on the CHSH expectation value $S$ and the one-body correlator $\avg{A_1}$. 
  The blue curve represents the boundary of the domain $\mathcal{D}\subset [0,1]\times[2,2\sqrt{2}]$ where the values of $(\avg{A_1},S)$ are consistent with quantum theory.
  We conjecture that the convex envelope of the function $\tilde g_q(\avg{A_1}, S) = g_{q} \bigro{\abs{\avg{A_{1}}}, \sqrt{S^2/4 - 1}}$ in $\mathcal{D}$ is obtained by taking a convex decomposition of the point $(1,2)$ and a point on the line from $(1,2)$ to $(\avg{A_1},S)$. The figure illustrates such a convex decomposition (red points) for the point $(0.5,2.2)$ (green point). From this, we can compute a candidate affine function \eqref{eq:tangent} that optimally certifies the entropy of the point $(0.5,2.2)$. Setting a value for $\varepsilon$, we then run a recursive algorithm to find a rectangle covering, depicted in the figure, that certifies the candidate affine function. We chose a value $\varepsilon = 0.025$ such that the resultant covering is coarse enough that it can be visualized, but much smaller values, e.g., $\varepsilon \approx 10^{-8}$ or less can readily be used.}
  \label{fig:recursive}
\end{figure}

An important difference with the linear programming approach above is that we do not necessarily have to decide on a covering $\mathcal{K}$ in advance. In fact, this is often very wasteful as, to obtain a good bound with a small tolerance, we would typically find we need a fine discretization of the domain only close to where the bound coincides with its tangent. Finding a suitable discretization can then be done naturally, and in practice often very rapidly, by starting by testing \eqref{eq:tangent-test} for the polytopes $K$ in an initially coarse covering (which could consist of just one polytope containing the entire domain) and then, for each $K$ for which the test fails, subdividing $K$ into smaller polytopes and recursively applying the test to each of those (see illustration in Figure~\ref{fig:recursive}).

\paragraph*{Application to the bound \eqref{eq:bias} including the bias $\avg{A_1}$.}
We used this recursive certification method, coupled with a guess on the optimal linear tradeoff functions, for the qubit bound \eqref{eq:bias} which depends on the two variables $\avg{A_1}$ and $S$. The function $\tilde{g}_q(\avg{A_1}, S) \equiv g_{q} \bigro{\abs{\avg{A_{1}}}, \sqrt{S^2/4 - 1}}$ defining this bound is not convex as its Hessian matrix is not positive semidefinite everywhere. It appears, though, to be convex in each of the parameters  $\avg{A_{1}}$ and $S$ individually, and more generally in any direction passing through the positive orthant in the plane $\avg{A_{1}}$--$S$. This implies that the convex envelope of $\tilde{g}_q(\avg{A_1},S)$ can be constructed by considering at most convex combinations of \emph{two} points in the plane, instead of three points as follows by Carath\'eodory's theorem. Indeed, any non-trivial convex combination of three points in the plane $\avg{A_{1}}$--$S$ would have at least two of those points joined by a segment aligned in the direction of the positive orthant. But since the function is convex in that direction, one can advantageously replace the two points by a mixture of those.

Furthermore, if we are interested in computing a valid entropy bound for a point with $\avg{A_1}$ positive, it is sufficient to consider convex combinations in the domain $\mathcal{D} \subset [0,1] \times [2,2\sqrt{2}]$ of the plane $\avg{A_{1}}$--$S$, i.e., points with negative values of $\avg{A_1}$ can be neglected. To see this, consider a convex combination
\begin{equation}
  (\avg{A_1}, S)
  = t \, \bigro{\avg{A_1}', S'} + (1-t) \, \bigro{\avg{A_1}'', S''}
\end{equation}
where $\avg{A_1}' < 0$ is negative for the point $(\avg{A_1}, S)$ yielding a corresponding value for the entropy function
\begin{equation}
  t \, \tilde{g}_q \bigro{\avg{A_1}', S'}
  + (1-t) \, \tilde{g}_q \bigro{\avg{A_1}'', S''}
\end{equation}
that is a valid lower bound for $H(A_1^q|E)$. Replace now this convex strategy by the (valid) convex combination
\begin{equation}
  (\avg{A_1}, S)
  = t \, (0, S') + (1-t) \, \biggro{\frac{\avg{A_1}}{1-t},\, S''} \,.
\end{equation}
The corresponding value for the entropy function is
\begin{equation}
  t \, \tilde{g}_q(0, S')
  + (1-t) \, \tilde{g}_q \biggro{\frac{\avg{A_1}}{1-t},\, S''} \,,
\end{equation}
which is still a valid lower bound for $H(A_1^q | E)$ because of the monotonicity property of the bound and the fact that $\frac{\avg{A_1}}{1-t} \leq \avg{A_1}''$ (since $\avg{A_1}' < 0$). 

Finally, we numerically observed that the convex envelope of $\tilde{g}_q(\avg{A_1}, S)$ in the domain $[0,1] \times [2,2\sqrt{2}]$ was always obtained by taking a convex decomposition of two particular points: the point $(1, 2)$ and a point on the line from $(1, 2)$ to $(\avg{A_1}, S)$. This observation gives a conjecture for the convex envelope of the qubit bound \eqref{eq:bias}, from which candidate linear tradeoff functions of the form \eqref{eq:tangent} can readily be computed as tangents to this envelope. We can then attempt to certify that such candidates are indeed proper tradeoff functions through a rectangle covering and the recursive procedure described above, as illustrated in Figure~\ref{fig:recursive}. We can in principle perform such certification to arbitrary precision $\varepsilon$, though, in practice, we may be limited by the number of rectangles required to reach a very small $\varepsilon$ and by the limited precision of hardware floating-point arithmetic on typical computers. The key rates and results presented in Section~\ref{sec:bias} have been computed using this procedure. From our results, it appears that our conjecture on the convex envelope of $\tilde g_q(\avg{A_1},S)$ is correct as we are always able to certify the resultant linear tradeoff functions up to a precision of the order of $\varepsilon \approx 10^{-6}$ or better.

\section{Applications}
\label{sec:applications}

Here, we apply our method to bound the asymptotic one-way key rate, given by the Devetak-Winter rate
\begin{equation}
  r = H(K_{A} | XE) - H(K_{A} | K_{B}) \,,
\end{equation}
for DIQKD in two situations of interest: white noise, where we assume that Alice and Bob share an attenuated version,
\begin{equation}
  \label{eq:noisy-phiplus}
  \rho = v \phi^{+} + (1 - v) \id / 4 \,,
\end{equation}
depending on some visibility $v$, of the ideal maximally-entangled state
\begin{equation}
  \ket{\phi^{+}} = \frac{1}{\sqrt{2}} \bigro{\ket{00} + \ket{11}} \,,
\end{equation}
and limited detection efficiency, where we assume that Alice's and Bob's devices return one of the expected outcomes $\pm 1$ with a probability $\eta$ less than one.

The qubit bound \eqref{eq:chsh-bound} (which is already convex) was used in \cite{ref:ab2007} to compute the key rate of the standard CHSH DIQKD protocol and the convexification of \eqref{eq:salpha-qubit-bound} was used in \cite{ref:wap2021} to generalize the analysis in terms of the asymmetric CHSH expressions $S_{\alpha}$ and incorporating noisy preprocessing. We will now illustrate the use of the two other qubit bounds \eqref{eq:bias} and \eqref{eq:HAXE-twobasis-qubit} given in the preceding section, in subsections~\ref{sec:bias} and \ref{sec:two-basis}, respectively.

In \cite{ref:wap2021}, the asymmetric CHSH expressions were chosen for parameter estimation because they retain the same symmetries as the version of the DIQKD protocol where only one of Alice's measurements, $A_{1}$, is used to generate the key and they can be used to derive the optimal one-way key rate for that protocol with respect to white noise. There is no analogous connection between the asymmetric CHSH expressions and losses and, in fact, the lowest threshold, $\eta \approx 82.57\%$, on the global detection efficiency reported in \cite{ref:wap2021} was obtained using CHSH (the special case of $S_{\alpha}$ with $\alpha = 1$).

In the following, we reanalyze these correlation models using different setups. In particular, as \cite{ref:wap2021} already does an optimal analysis for white noise using one measurement basis for key generation and with noisy preprocessing, the only remaining way to improve the noise robustness is to use a different protocol. For that case, we apply our approach to a variant of the protocol based on CHSH, proposed recently in \cite{ref:sg2021}, in which both of Alice's measurements $A_{1}$ and $A_{2}$ are used to generate the key. For losses, by contrast, as remarked in \cite{ref:wap2021} the analysis performed there was likely not optimal as the treatment of losses introduced biases in the probabilities of Alice's and Bob's measurement outcomes, while the analytic bound on the entropy used there was optimized for the case that Alice's outcomes are obtained equiprobably. For losses, therefore, we concentrate on bounding the key rate using the expectation value $\avg{A_{1}}$ of Alice's key-generation measurement in addition to the Bell violation.

\subsection{White noise analysis for the two-basis protocol}
\label{sec:two-basis}

In the two-basis protocol of \cite{ref:sg2021}, Alice and Bob ideally share a maximally-entangled state $\ket{\phi^{+}}$ and have devices that, for Alice, ideally perform the two measurements
\begin{IEEEeqnarray}{c+c}
  A_{1} = \sz \,, & A_{2} = \sx \,,
\end{IEEEeqnarray}
and, for Bob, the four measurements
\begin{IEEEeqnarray}{rCl+rCl}
  B_{1} &=& \frac{\sz + \sx}{\sqrt{2}} \,, & B_{3} &=& \sz \,, \\
  B_{2} &=& \frac{\sz - \sx}{\sqrt{2}} \,, & B_{4} &=& \sx \,.
\end{IEEEeqnarray}
This ideal realization is designed so that the measurements $A_{1}$, $A_{2}$, $B_{1}$, and $B_{2}$ yield a maximal violation of the CHSH Bell inequality while Bob's measurements $B_{3}$ and $B_{4}$ yield outcomes that are perfectly correlated with Alice's when she measures, respectively, $A_{1}$ and $A_{2}$, i.e., $\avg{A_{1} B_{3}} = \avg{A_{2} B_{4}} = 1$.

In the protocol, Alice and Bob use rounds where Bob measures $B_{1}$ or $B_{2}$ to estimate CHSH; they use a small fraction of the rounds where Bob measures $B_{3}$ and $B_{4}$ to estimate how correlated the outcomes are with $A_{1}$ and $A_{2}$, and use the results of the remaining rounds where Alice and Bob measured $A_{1}$ and $B_{3}$ or $A_{2}$ and $B_{4}$ as their raw key. We also assume in the following that Alice flips her outcomes in the key generation rounds (i.e., applies noisy preprocessing) with some probability $q$.

Let us suppose that Alice uses the measurements $A_{1}$ and $A_{2}$ with probabilities $p'$ and $\bar{p}' = 1 - p'$ and that Bob uses the measurements $B_{3}$ and $B_{4}$ with the same relative probabilities. Then, out of the rounds not used for parameter estimation, the asymptotic key rate, taking into account the effect of sifting\footnote{In particular, the key rate is attenuated by the probability $p^{\prime 2} + \bar{p}^{\prime 2}$ that Alice and Bob use matching bases. It has been pointed out in \cite{ref:ts2020} that this can be avoided, but this requires the parties to either possess quantum memories or to use a very long preshared key to coordinate the measurement choices.}, is
\begin{IEEEeqnarray}{rCl}
  r &=& p^{\prime 2} \, r_{13} + \bar{p}^{\prime 2} \, r_{24} \nonumber \\
  &=& (p^{\prime 2} + \bar{p}^{\prime 2})
  (p \, r_{13} + \bar{p} \, r_{24}) \,, 
\end{IEEEeqnarray}
where
\begin{equation}
  r_{xy} = H(A^q_{x} | E) - H(A^q_{x} | B_{y})
\end{equation}
and we introduced $p = {p'}^{2} / ({p'}^{2} + \bar{p}^{\prime 2})$ and $\bar{p} = 1 - p$ in the second line. Here, $H(A^q_{1} | B_{3})$ and $H(A^q_{2} | B_{4})$ depend only on the correlations between Alice's and Bob's measurement outcomes, which they know from parameter estimation. Assuming Alice and Bob perform the ideal measurements on an attenuated state \eqref{eq:noisy-phiplus}, the entropies of Alice's outcomes conditioned on Bob are
\begin{equation}
  H(A^q_1 | B_3) = H(A^q_2 | B_4) = h \bigro{q + \delta (1-2q)} \,,
\end{equation}
where the channel error rate $\delta$ is related to the visibility $v$ in \eqref{eq:noisy-phiplus} by $v = 1 - 2 \delta$, while the CHSH expectation value is
\begin{equation}
  S = 2 \sqrt{2} (1 - 2 \delta) \,.
\end{equation}

Bounding the key rate thus amounts to establishing a lower bound on the weighted average conditional entropy
\begin{equation}
  p H(A^q_{1} | E) + \bar{p} H(A^q_{2} | E) = H(A_{X}^q | X E)
\end{equation}
depending on the CHSH violation. A valid \emph{qubit} bound in terms of the CHSH expectation value $S$ is given by \eqref{eq:HAXE-twobasis-qubit}, from which a valid, fully device-independent, convex lower bound can be obtained using the techniques discussed in Section~\ref{sec:linear-programming}.

We can thus express the bound we obtain on the key rate, via CHSH, in terms of $\delta$ using our approach as
\begin{equation}
  r \geq (p^{\prime 2} + \bar{p}^{\prime 2}) \Bigsq{
    \tilde{f}_{q} \bigro{2\sqrt{2}(1-2\delta)}
    - h \bigro{q + \delta (1-2q)}} \,,
\end{equation}
where $\tilde{f}_{q}(S)$ is the convex lower bound we obtain for the entropy, evaluated at $S = 2\sqrt{2}(1 - 2 \delta)$.

We remark here that we could, in principle, bound the average entropy in terms of any correlation Bell inequality. We use only the CHSH expectation value here both for simplicity and because, in the most interesting case where the bases are used equiprobably (i.e., $p = 1/2$), we can infer from the symmetries of the protocol that CHSH is already the optimal measure of nonlocality for white noise (see Appendix~\ref{ap:twobasis-chsh} for details).

\begin{figure}[tbp]
  \includegraphics{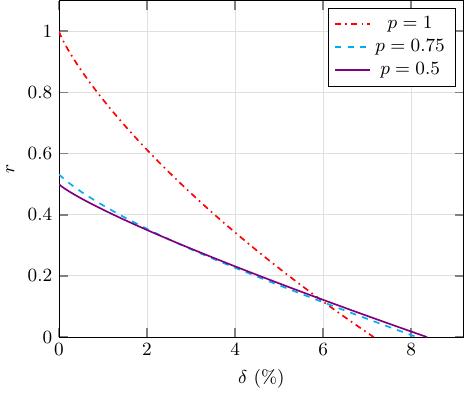}
  \caption{Lower bound on the Devetak-Winter rate as a function of the channel error rate $\delta$, assuming $q=0$.}\label{fig:two-bas}
\end{figure}

\begin{figure}[tbp]
  \includegraphics{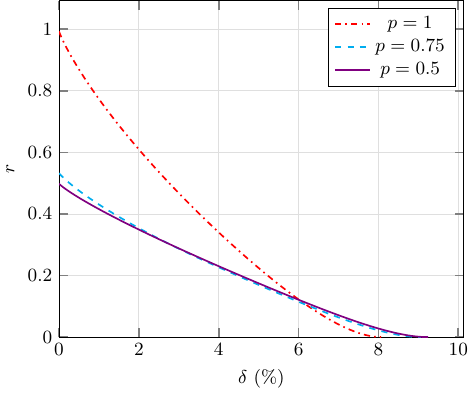}
  \caption{Lower bound on the Devetak-Winter rate as a function of the channel error rate $\delta$, using an optimal noisy preprocessing.}\label{fig:two-bas-qopt}
\end{figure}

The key rate we obtain using our approach for $p = 0.5$ and $p = 0.75$ are illustrated, and compared with the known analytical bounds for $p = 1$, without noisy preprocessing (i.e., $q = 0$) and with the optimal amount of noisy preprocessing applied in Figures~\ref{fig:two-bas} and \ref{fig:two-bas-qopt}. The threshold noise rates up to which we obtain a positive key rate are reported for different values of $q$ in Table~\ref{tab:two-basis}. For $q = 0$ and $q$ close to $1/2$, the results essentially rigorously confirm the thresholds of $8.36\%$ and $9.24\%$ that were anticipated could be obtained in the conclusion of \cite{ref:wap2021}. For $0 < p < 1/2$, similar to \cite{ref:sg2021}, we did not see any improvement to the key rate; the highest rate appeared to always be obtained with either $p = 1$ or $p = 1/2$, depending on the value of $S$. However, as it may not be realistic to be sure that the measurements are used \emph{exactly} equiprobably in a real implementation, we note that it is important to be able to bound the entropy for values of $p$ that may deviate a little from $0.5$. The key rate is in fact very robust against deviations of $p$ from $0.5$, as can be seen comparing the results for $p = 0.5$ and $p = 0.75$ in Figures~\ref{fig:two-bas} and \ref{fig:two-bas-qopt}.

The best threshold of $9.24\%$ obtained for $q$ close to 1/2 using our method is close to the best threshold of $9.33\%$  recently reported in \cite{ref:ts2020} and obtained for $q = 0.3$, although the method we have used allows the key rate to be bounded much more rapidly\footnote{Ref.~\cite{ref:ts2020} reports requiring $\sim 5000$ processor-core hours to obtain a numerical bound on the average conditional entropy. For comparison, using our method we could generate a plot of the conditional entropy with 500 points in a minute or two on a regular laptop using the Lasserre hierarchy or almost instantaneously using the analytic method for $p = 1/2$ described in Appendix~\ref{ap:two-basis}.}. Without noisy preprocessing, the threshold of $8.36\%$ we obtain is slightly better than the threshold around $8.24\%$ found in \cite{ref:sg2021} and the same as the threshold that would be obtained using the ``conjectured alternative proof'' (after taking the convex envelope of the result) proposed in section~I.H of the supplementary information to the same paper\footnote{This is not a coincidence. The section in question proposes to bound the key rate using a lower bound on the conditional entropy in terms of the fidelity of Eve's marginal states. This is very closely related to the BB84 bound \cite{ref:w2013} and, in fact, all of the lower bounds we derive on the correlation terms $\abs{\avg{\bar{A}_{x} \otimes B}}$ appearing in the BB84 bounds we use are also (typically tight) lower bounds on the fidelity of Eve's marginals following the qubit reduction.}.

\begin{table}[tbp]
  \centering\footnotesize
  \begin{tabular}{c | c c c c c}
    $p$               & $q=0$ & $q=0.2$ & $q=0.3$ & $q=0.49$  & $q \to 1/2$ \\
    \hline
    1  & 7.1492 & 7.9503 & 8.0321 & 8.0848  & 8.0848\\
    0.5   & 8.3599 & 9.1130 & 9.1923& 9.2434 & 9.2435 \\
  \end{tabular}
  \caption{Threshold error rates ($\%$) obtained for different probabilities $p$ of measuring $A_1$ after sifting non-matching basis.} \label{tab:two-basis}
\end{table}\normalsize

We provide an indication of how close the key-rate bound we obtain in the case $p=1/2$ is to being optimal by comparing with a specific strategy, which was already identified as a likely candidate for the optimal collective attack for $q = 0$ in \cite{ref:wap2021}, and described in Appendix~\ref{sec:attack-two-basis}. This attack yields the following value for the average entropy
\begin{equation}
  \label{eq:twobasis-conjecture}
  \tfrac{1}{2} H(A^q_{1} | E) + \tfrac{1}{2} H(A^q_{2} | E)
  = \bar{f}_{q} \bigro{S / \sqrt{8}} \,,
\end{equation}
where
\begin{equation}
  \bar{f}_{q}(x) = \begin{cases}
    f_{q}(x) &\text{if } x \geq x_{*} \\
    h(q) + f'_{q}(x_{*}) (x - 1/\sqrt{2}) &\text{if } x \leq x_{*}
  \end{cases}
\end{equation}
with $x_{*}$ (dependent on $q$) such that
\begin{equation}
  h(q) + f'_{q}(x_{*}) (x - 1/\sqrt{2}) = f_{q}(x_{*}) \,,
\end{equation}
and where $f_q(x)$ is defined in Eq.~\eqref{eq:fqx}.

\begin{figure}[htbp]
  \includegraphics{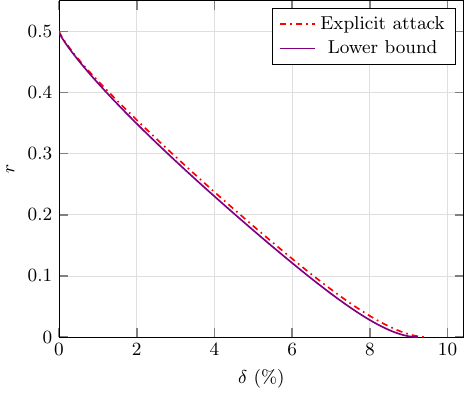}
  \caption{Comparison between the conjectured optimal attack and the lower bound on the Devetak-Winter rate as a function of the channel error rate $\delta$, using an optimal noisy preprocessing.}\label{fig:two-bas-conj}
\end{figure}

\begin{figure}[htbp]
  \includegraphics{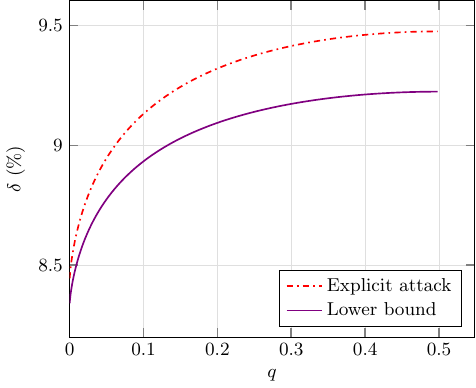}
  \caption{Thresholds for the channel error rate as a function of the noisy preprocessing computed using the conjectured optimal attack and our lower bound on the conditional entropy.}\label{fig:two-bas-thr}
\end{figure}

The results of numerical tests done without noisy preprocessing in \cite{ref:wap2021} and \cite{ref:brc2021} strongly suggest that \eqref{eq:twobasis-conjecture} actually gives the optimal bound on the average entropy for $q = 0$. Additional tests we did for this work did not find a counterexample for $q \neq 0$. But even without a proof of optimality, as \eqref{eq:twobasis-conjecture} is obtained with a known collective attack it gives an upper bound on the one-way asymptotic key rate with noisy preprocessing. A comparison of the key rates, optimized over $q$, using our numerical lower bound (already given in Figure~\ref{fig:two-bas-qopt}) and using \eqref{eq:twobasis-conjecture} is given in Figure~\ref{fig:two-bas-conj} and shows the two to be very close. The threshold error rate obtained using \eqref{eq:twobasis-conjecture} ranges from $\delta \approx 8.4447\%$ for $q = 0$ up to $\delta \approx 9.4756\%$ for $q \to 1/2$, and is compared with the threshold obtained using our numerical method in Figure~\ref{fig:two-bas-thr}.

\subsection{More refined loss analysis exploiting bias}
\label{sec:bias}

Here, we consider a setup where we suppose that the main imperfection is that Alice's and Bob's devices have a detection efficiency that is less than perfect, i.e., we suppose that, in each protocol round, each of their devices outputs one of the regular outcomes $\pm 1$ with probability $\eta$ and outputs nothing, or a ``nondetection'' outcome $\emptyset$, with probability $1 - \eta$. In order to use our approach, which strictly applies to protocols in which the measurements in the Bell test have binary outcomes, we map nondetection events resulting from the measurements $A_{1}$, $A_{2}$, $B_{1}$, and $B_{2}$ used to perform the Bell test to $+1$.

In this case we consider the usual, single-basis, version of the DIQKD protocol, but with different states and measurements. Similar to the Eberhard scheme \cite{ref:e1993}, we suppose that Alice and Bob (ideally) share a partially-entangled two-qubit state
\begin{equation}
  \ket{\psi_\theta} = \cos \bigro{\tfrac{\theta}{2}} \ket{00}
  + \sin \bigro{\tfrac{\theta}{2}} \ket{11} \,,
\end{equation}
and that Alice and Bob (ideally) perform, respectively, two and three measurements
\begin{IEEEeqnarray}{rCl+rCl}
  A_{x} &=& \cos(\varphi_{A,x}) \sz + \sin(\varphi_{A,x}) \sx \,,
  & x &=& 1,2 \\
  B_{y} &=& \cos(\varphi_{B,y}) \sz + \sin(\varphi_{B,y}) \sx \,,
  & y &=& 1,2,3 \,, \IEEEeqnarraynumspace
\end{IEEEeqnarray}
determined by angles $\varphi_{A,x}$ and $\varphi_{B,y}$ that we will optimize over when bounding the key rate\footnote{Note that this is a slight generalization with respect to \cite{ref:wap2021}, which fixed $A_{1}$ and $B_{3}$ to $\sz$.}. Alice and Bob use the measurements $A_{1}$, $A_{2}$, $B_{1}$, and $B_{2}$ to estimate the CHSH expectation value and use $A_{1}$ and $B_{3}$ to generate the key.

As we are only considering the usual single-basis version of the protocol, the asymptotic key rate is
\begin{equation}
  r = H(A^q_{1} | E) - H(A^q_{1} | B_{3})
\end{equation}
where the Shannon entropy of Alice's outcome conditioned on Bob,
\begin{equation}
  H(A^q_1 | B_3) = -\sum_{a,b} p(a,b) \log_2 \bigro{p(a|b)} \,,
\end{equation}
depends on the joint probability $p(a,b)$ that Alice obtains the outcome $a \in \{+1, -1\}$ from measuring $A_{1}$ after mapping nondetection events to $+1$ and flipping the result with probability $q$, and Bob obtains the outcome $b \in \{+1, -1, \emptyset\}$ from measuring $B_{3}$ and possibly obtaining the loss outcome $\emptyset$ with probability $1 - \eta$.

To bound the key rate we need to bound $H(A^q_{1} | E)$. As mentioned above, mapping nondetection events deterministically to $+1$ and deliberately using a partially-entangled state bias Alice's and Bob's measurements to giving one of the outcomes more frequently than the other. We can exploit this by taking into account the expectation value $\avg{A_{1}}$ of Alice's key generation measurement, in addition to the CHSH expectation value $S$, to derive a better lower bound on the entropy. 

\begin{figure}[tbp]
  \includegraphics[scale=0.82]{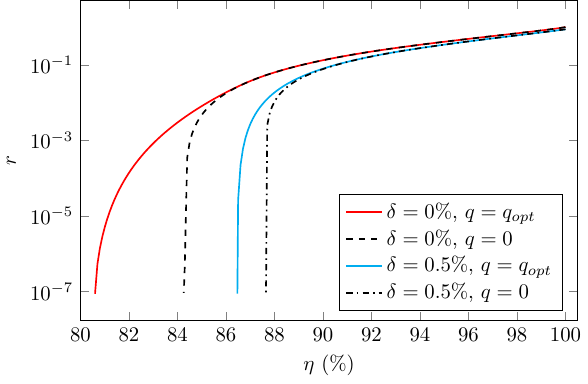}
  \caption{Key rate as a function of the detection efficiency with no channel error rate and with a little error rate.}
  \label{fig:eta-rate}
\end{figure}

The expectation value $\avg{A_{1}}$ can be taken into account using the qubit bound \eqref{eq:bias} and the convexification procedure discussed at the end of Section~\ref{sec:affine-tradeoff} and illustrated in Figure~\ref{fig:recursive}.  Using this approach, we optimized the key rate numerically over the angles $\varphi_{A_j}$, $\varphi_{B_k}$, and $\theta$. 
The optimized key rates, both assuming no noise and a white noise rate of $\delta = 0.5\%$ are represented both for $q = 0$ and with optimized $q$ in Figure~\ref{fig:eta-rate}. 

\begin{figure}[tbp]
  \includegraphics{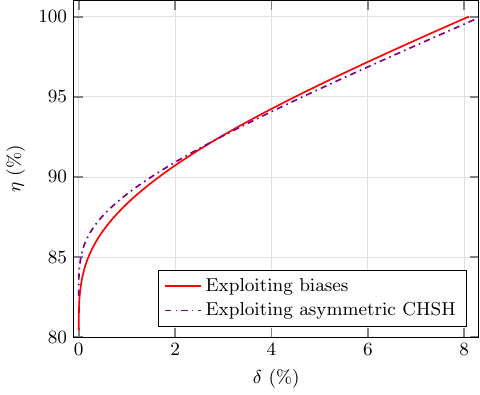}
  \caption{Threshold detection efficiency $\eta$ as a function of the channel error rate $\delta$.}
  \label{fig:eta-delta}
\end{figure}

As one can see in the figure, the highest key rate is very small for a significant range of global detection efficiencies close to the threshold as a result of being obtained for values of $q$ close to 1/2 and very weakly entangled states. Due to this, the threshold detector efficiency above which a positive key rate can be certified is very sensitive and, for example, significantly worsened by the addition of even a small amount of depolarizing noise. To illustrate this, we plot the threshold global detection efficiency as a function of the error rate $\delta$ in Figure~\ref{fig:eta-delta}, where a comparison is provided with the earlier results of \cite{ref:wap2021} using the analytic entropy bound for the asymmetric CHSH expressions.

Table~\ref{tab:no-error} gives the thresholds on the detection efficiency that we find using our approach for different values of $q$ assuming no additional noise. We include in the table both the thresholds for which we can certify a positive key rate and the ones obtained using our conjecture regarding the convex envelope of the qubit bound. The small discrepancy between the two values, particularly for larger values of $q$, is due to the difficulty of numerically certifying the key rate accurately when the key rate becomes very small (the key rate for the last column of Table~\ref{tab:no-error} is of $O(10^{-12})$). Indeed to certify the entropy to a very high precision using a discretized qubit bound requires using a very dense covering, which at some point becomes too time-consuming computationally. 

This issue however only affects the certification of extremely small asymptotic key rates, such as the long tail observed in Figure~\ref{fig:eta-rate}, which are probably too low to be of practical value and likely to be dwarfed by the difference made by even small amounts of noise or corrections due to finite-key effects. To illustrate this, in Table~\ref{tab:error} we report the detection efficiency thresholds in the presence of a channel noise rate of $\delta = 0.5$. In this case, the thresholds using the conjectured convex envelope and those that can be properly certified are the same up to the precision to which we report the results.

\begin{table}[tbp]
  \centering\footnotesize
  \begin{tabular}{c | c c c c}
                   & $q=0$ & $q=0.2$ & $q=0.3$ & $q=0.49$  \\
    \hline
    Certified  & 84.2149 & 80.4642 & 80.3411  & 80.2593  \\
  Conjectured   & 84.2147 & 80.4362 & 80.3046  & 80.2283  \\
  \end{tabular}
  \caption{Threshold detection efficiencies ($\%$) for different probabilities $q$ of flipping Alice's outcome assuming no channel noise. For $q$ between $0.49$ and $0.5$, we did not observe an improvement of the threshold up to the precision reported in the table.}
  \label{tab:no-error}
\end{table}

\begin{table}[tbp]
  \centering\footnotesize
  \begin{tabular}{c | c c c c}
              & $q=0$ & $q=0.2$ & $q=0.3$ & $q=0.49$  \\
    \hline
   $\delta=0.5\%$  & 87.6017 & 86.5842 & 86.5013 & 86.4490 \\
  \end{tabular}
  \caption{Certified threshold detection efficiencies ($\%$) obtained for different probabilities $q$ of flipping Alice's outcome and with $\delta=0.5\%$ of channel error rate. We do not observe a difference with the conjectured case up to the precision reported in the table.}
  \label{tab:error}
\end{table}

Finally, we remark that the qubit bound \eqref{eq:bias} is tight in $\avg{A_{1}}$ and $S$ for all $q$ as there is an explicit attack, described in Appendix~\ref{sec:attack-bias}, that saturates it. This means that our conjecture regarding the convex envelope of the qubit bound represents a valid attack yielding upper bounds on the key rate (as it corresponds to an explicit mixture of two-qubit strategies). This means that the certified bounds that we report in Table~\ref{tab:error} are, up to the precision we use, optimal in terms of $\avg{A_1}$ and $S$, and that the second line of Table~\ref{tab:no-error} corresponds to the minimal detection thresholds one can hope to attain using only information about $\avg{A_1}$ and $S$.

\section{Discussion}

Building on \cite{ref:wap2021}, we have introduced a flexible approach to derive practical and fully device-independent bounds on the key rate for DIQKD in the 2-input/2-output setting. We have illustrated it on to the two-basis variant of the CHSH DIQKD protocol as well as to undertake a more optimized analysis of the single-basis variant when the main anticipated experimental imperfection is losses. Contrarily to \cite{ref:wap2021}, we used numerical methods to solve part of the problem in both cases and obtain optimal or close to optimal bounds on the conditional entropy within a very low amount of computation time. The results may be used to derive bounds on the key rate in the asymptotic limit or in the finite-key regime via the entropy accumulation theorem. They may also be useful as a point of comparison with different numerical approaches used to bound the conditional entropy in the device-independent setting.

When considering losses we found that the global detection efficiency can be brought under $80.26\%$. This is notably below the detection efficiency of $87.49\%$ attained in the recent experimental demonstration of device-independent quantum key distribution based on a photonic setup \cite{ref:liu2022toward}. As we remarked in the previous section, however, our threshold is attained using a very weakly entangled state and increases significantly if any realistic amount of noise is added to the model we studied. (Separately, a finite-key analysis would likely have the same effect.)

While writing this manuscript, a new promising numerical method to bound the conditional entropy in general DI scenarios was proposed \cite{ref:bff2021b}. Our detection threshold, derived using only the expectation value $\avg{A_{1}}$ of Alice's key-generation measurement in addition to CHSH, is slightly lower than the threshold of $80.5\%$ reported in \cite{ref:bff2021b} using full statistics. This is not a limitation of the method of \cite{ref:bff2021b}, but rather a matter of using a suboptimal state and measurement implementation parameters in that work. Indeed, running their method on the correlations achieving the threshold of $80.2593\%$ in Table~\ref{tab:no-error}, the authors of \cite{ref:bff2021b} confirmed to us that they also find a positive key rate \cite{ref:b2021} (though, again, using full statistics instead of only $\avg{A_1}$ and $S$). This illustrates the interest of having complementary methods. While \cite{ref:bff2021b} can in principle be used to tackle very general problems, our method specializing on the 2-input/2-output scenario allows us to rapidly explore the parameter space to find a good implementation. Moreover, there exist scenarios in which our analysis can provide slightly better bounds compared to the numerical method as one can observe from \cite[Figure 6b]{ref:bff2021b}.

A recent result \cite{ref:zhang2021quantum} obtained lower bounds on the key rate for the finite-size case without the use of the entropy accumulation theorem in the two-input/two-output scenario. It might be interesting to investigate whether our results involving different parameters to bound the conditional Von Neumann entropy can be used in combination with their technique.

Finally, although we discussed in detail two specific examples illustrating our approach to bounding the conditional von Neumann entropy, we point out that other bounds can be derived. For instance, we could combine the BB84-type bound \eqref{eq:HZE-A1-XB} using bias with the correlation bound \eqref{eq:Salpha-XB-bound} in terms of the asymmetric CHSH expectations. As suggested by Figure~\ref{fig:eta-delta}, this should slightly improve the analysis presented here (are least for larger amounts of noise $\delta$). One could also, much more generally, use numerical techniques \cite{ref:wlc2018} to derive device-dependent bounds on the conditional von Neumann entropy that are more stringent and combine them with correlation bounds involving full-statistics obtained through relaxations of the Lasserre hierarchy. Our method can also in principle be applied to the $n$-partite setting, e.g., to derive entropy bounds based on Mermin-type Bell inequalities \cite{ref:m1990,ref:gm2021}.

The code used to obtain the numerical results in this paper is available on GitHub \cite{ref:m2021}.

\phantom{}

\begin{acknowledgments}
  This work was supported by the EU Quantum Flagship project QRANGE\@ and the F.R.S-FNRS through the grant PDR T.0171.22. S.P. is a Senior Research Associate of the Fonds de la Recherche Scientifique -- FNRS\@.
\end{acknowledgments}

\bibliographystyle{unsrtnat}
\bibliography{references}

\clearpage
\appendix

\section{Derivation of BB84 bound with bias}
\label{ap:part-sym}

The BB84 entropy bound \eqref{eq:HZE-A1-XB} is a generalization of the two bounds \eqref{eq:bb84-bound} and \eqref{eq:HZE-bb84-noise}, which give the special cases of \eqref{eq:HZE-A1-XB} with $\avg{A_{1}} = 0$ and both with $\avg{A_{1}} = 0$ and no noisy preprocessing ($q = 0$). It can be derived, in a way that also confirms the monotonicity property, essentially by modifying the symmetrization step in the derivation done in section~4.2 of the
paper \cite{ref:wap2021}. We do this in detail here.

As in the derivation of \cite{ref:wap2021}, we suppose that Alice, Bob, and Eve share a pure tripartite state
\begin{equation}
  \ket{\psi}_{ABE} = \ket{0}_{A} \ket{\psi_{0}}_{BE}
  + \ket{1}_{A} \ket{\psi_{1}}_{BE} \,,
\end{equation}
where $\ket{0}$ and $\ket{1}$ are the eigenstates of $A_{1}$, which we identify here with $\sz$, and $\ket{\psi_{0}}$ and $\ket{\psi_{1}}$ are arbitrary (and not necessarily orthogonal) states shared by Bob and Eve normalized so that
\begin{equation}
  \braket{\psi_{0}}{\psi_{0}} + \braket{\psi_{1}}{\psi_{1}} = 1 \,.
\end{equation}
After Alice measures $A_{1} = \sz$ and flips the outcome with probability $q$, the correlations between Alice and Eve are described by the classical-quantum state
\begin{equation}
  \tau_{AE}
  = [0]_{A} \otimes (\bar{q} \psi^{E}_{0} + q \psi^{E}_{1})
  + [1]_{A} \otimes (q \psi^{E}_{0} + \bar{q} \psi^{E}_{1}) \,,
\end{equation}
where $\bar{q} = 1 - q$ and $\psi^{E}_{a} = \Tr_{B}[\psi_{a}]$ are the partial traces of the states $\ket{\psi_{a}}$ accessible to Eve.

Now, since renaming the outcomes does not change the entropy, the conditional entropy $H(\sz | E) = H(\sz E) - H(E)$ computed on the above state is the same as the conditional entropy computed on
\begin{equation}
  \tau'_{AE}
  = [1]_{A} \otimes (\bar{q} \psi^{E}_{0} + q \psi^{E}_{1})
  + [0]_{A} \otimes (q \psi^{E}_{0} + \bar{q} \psi^{E}_{1}) \,,
\end{equation}
which is the same state as above except that we have swapped $[0]_{A}$ and $[1]_{A}$. They in addition have the same entropy as a partly symmetrized state,
\begin{equation}
  \bar{\tau}_{AEF}
  = \bar{p} \, \tau_{AE} \otimes [0]_{F}
  + p \, \tau'_{AE} \otimes [1]_{F} \,,
\end{equation}
for any probability $p$ and $\bar{p} = 1 - p$, since
\begin{equation}
  H(\sz | EF)_{\bar{\tau}}
  = \bar{p} \, H(\sz | E)_{\tau} + p \, H(\sz | E)_{\tau'}
  = H(\sz | E)_{\tau} \,.
\end{equation}
The above state, written out explicitly, is
\begin{IEEEeqnarray}{rCll}
  \bar{\tau}_{AEF} &=& \IEEEeqnarraymulticol{2}{l}{
    \begin{IEEEeqnarraybox}[][t]{rl}
      [0]_{A} \otimes \Bigsq{ &
        \bar{p} (\bar{q} \psi^{E}_{0} + q \psi^{E}_{1}) \otimes [0]_{F} \\
        &+\> p (q \psi^{E}_{0} + \bar{q} \psi^{E}_{1}) \otimes [1]_{F}}
    \end{IEEEeqnarraybox}} \nonumber \\
  &&+\> [1]_{A} \otimes \Bigsq{ &
    \bar{p} (q \psi^{E}_{0} + \bar{q} \psi^{E}_{1})
    \otimes [0]_{F} \nonumber \\
    &&&+\> p (\bar{q} \psi^{E}_{0} + q \psi^{E}_{1})
    \otimes [1]_{F}} \,. \IEEEeqnarraynumspace
\end{IEEEeqnarray}
We rewrite this as
\begin{equation}
  \bar{\tau}_{AEF}
  = [0]_{A} \otimes (\bar{q} \sigma_{=} + q \sigma_{\neq})
  + [1]_{A} \otimes (q \sigma_{=} + \bar{q} \sigma_{\neq})
\end{equation}
with the (unnormalized) states
\begin{IEEEeqnarray}{rCl}
  \sigma_{=} &=& \bar{p} \, \psi^{E}_{0} \otimes [0]_{F}
  + p \, \psi^{E}_{1} \otimes [1]_{F} \,, \\
  \sigma_{\neq} &=& \bar{p} \, \psi^{E}_{1} \otimes [0]_{F}
  + p \, \psi^{E}_{0} \otimes [1]_{F} \,.
\end{IEEEeqnarray}
The state can be obtained as the marginal of an extended one,
\begin{IEEEeqnarray}{rCl}
  \label{eq:tauABEEFF}
  \bar{\tau}_{ABEE'FF'}
  &=& [0]_{A} \otimes (\bar{q} \chi_{=} + q \chi_{\neq}) \nonumber \\
  &&+\> [1]_{A} \otimes (q \chi_{=} + \bar{q} \chi_{\neq}) \,,
\end{IEEEeqnarray}
where $\ket{\chi_{=}}, \ket{\chi_{\neq}} \in \hilb{B} \otimes \hilb{E} \otimes \hilb{E'}
\otimes \hilb{F} \otimes \hilb{F'}$ are unnormalized pure states
\begin{IEEEeqnarray}{rCl}
  \ket{\chi_{=}} &=& \sqrt{\bar{p}} \ket{\psi_{0}} \ket{\phi_{0}} \ket{00}
  + \sqrt{p} \ket{\psi'_{1}} \ket{\phi_{1}} \ket{11} \,, \\
  \ket{\chi_{\neq}}
  &=& \sqrt{\bar{p}} \ket{\psi'_{1}} \ket{\phi_{1}} \ket{00}
  + \sqrt{p} \ket{\psi_{0}} \ket{\phi_{0}} \ket{11} \,,
\end{IEEEeqnarray}
in which
\begin{equation}
  \ket{\psi'_{1}} = e^{i\varphi} B \otimes \id_{E} \ket{\psi_{1}}
  \in \hilb{B} \otimes \hilb{E} \,,
\end{equation}
where $B$ is a Hermitian unitary operator (thus satisfying $B^{2} = \id_{B}$) acting on $\hilb{B}$ and $\varphi$ is a phase chosen such that $\braket{\psi_{0}}{\psi'_{1}}$ is real and nonnegative, and
\begin{equation}
  \ket{\phi_{0}}, \ket{\phi_{1}} \in \hilb{E'}
\end{equation}
are normalized states chosen to have some nonnegative real overlap
$\braket{\phi_{0}}{\phi_{1}} = \lambda_{\sx} \in [0, 1]$.

Using that the conditional entropy cannot increase if we extend the Hilbert
space being conditioned on, direct calculation of the conditional entropy on
the state \eqref{eq:tauABEEFF} gives
\begin{IEEEeqnarray}{rCl}
  H(\sz | E)_{\tau} &=& H(\sz | EF)_{\bar{\tau}} \nonumber \\
  &\geq& H(\sz | BEE'FF')_{\bar{\tau}} \nonumber \\
  &=& S(\bar{\tau}_{ABEE'FF'}) - S(\chi_{=} + \chi_{\neq}) \nonumber \\
  &=& H(\vect{\lambda}) - \phi \Bigro{\sqrt{Z'^{2} + X'^{2}}} \,,
\end{IEEEeqnarray}
where
\begin{IEEEeqnarray}{rCl}
  Z' &=& \norm{\chi_{=}} - \norm{\chi_{\neq}}
  \equiv \snorm{\chi_{=}} - \snorm{\chi_{\neq}} \,, \\
  X' &=& 2 \abs{\braket{\chi_{=}}{\chi_{\neq}}} \,,
\end{IEEEeqnarray}
and $H(\vect{\lambda}) = - \sum_{jk} \lambda_{jk} \log_{2}(\lambda_{jk})$ is
the Shannon entropy associated to the four eigenvalues of
\eqref{eq:tauABEEFF},
\begin{IEEEeqnarray}{rCl}
  \lambda_{11}
  &=& \frac{1}{4} \Bigsq{1 + Q Z' + \sqrt{R'^{2} + 2 Q Z'}} \,, \\
  \lambda_{12}
  &=& \frac{1}{4} \Bigsq{1 - Q Z' + \sqrt{R'^{2} - 2 Q Z'}} \,, \\
  \lambda_{21}
  &=& \frac{1}{4} \Bigsq{1 - Q Z' - \sqrt{R'^{2} - 2 Q Z'}} \,, \\
  \lambda_{22}
  &=& \frac{1}{4} \Bigsq{1 + Q Z' - \sqrt{R'^{2} + 2 Q Z'}} \,,
\end{IEEEeqnarray}
where $Q$ is related to the amount of noisy preprocessing applied by
\begin{equation}
  Q = \bar{q} - q = 1 - 2 q
\end{equation}
and
\begin{equation}
  R' = \sqrt{Z'^{2} + Q^{2} +  (1 - Q^{2}) X'^{2}} \,.
\end{equation}
We can factorize the four eigenvalues above as $\lambda_{jk} = p_{j} p'_{k}$
with
\begin{IEEEeqnarray}{rCl}
  p_{1} &=& \frac{1}{2} + \frac{1}{4} \bigro{R'_{+} + R'_{-}} \,, \\
  p_{2} &=& \frac{1}{2} - \frac{1}{4} \bigro{R'_{+} + R'_{-}} \,, \\
  p'_{1} &=& \frac{1}{2} + \frac{1}{4} \bigro{R'_{+} - R'_{-}} \,, \\
  p'_{2} &=& \frac{1}{2} - \frac{1}{4} \bigro{R'_{+} - R'_{-}} \,,
\end{IEEEeqnarray}
and
\begin{equation}
  R'_{\pm} = \sqrt{R'^{2} \pm 2 Q Z'} \,,
\end{equation}
so that $H(\vect{\lambda}) = H(\vect{p}) + H(\vect{p}')$. This allows us to
express the qubit entropy bound more concisely as
\begin{equation}
  H(\sz | E) \geq g_{q}(Z', X')
\end{equation}
with
\begin{IEEEeqnarray}{rCl}
  g_{q}(Z', X') &=& \phi \bigro{\tfrac{1}{2} (R'_{+} + R'_{-})}
  + \phi \bigro{\tfrac{1}{2} (R'_{+} - R'_{-})} \nonumber \\
  &&-\> \phi \bigro{\sqrt{Z'^{2} + X'^{2}}}
\end{IEEEeqnarray}
and
\begin{equation}
  R'_{\pm} = \sqrt{(Q \pm Z')^{2} + (1 - Q^{2}) X'^{2}} \,.
\end{equation}

At this point, we have recovered the form of the function $g_{q}$ defined in
section~\ref{sec:approach}. To complete the derivation note that, from the
definitions of $\ket{\chi_{=}}$ and $\ket{\chi_{\neq}}$ we have
\begin{IEEEeqnarray}{rCl}
  Z' &=& \norm{\chi_{0}} - \norm{\chi_{1}} \nonumber \\
  &=& \bar{p} \norm{\psi_{0}} + p \norm{\psi_{1}}
  - \bar{p} \norm{\psi_{1}} - p \norm{\psi_{0}} \nonumber \\
  &=& \lambda_{\sz} \bigro{\norm{\psi_{0}} - \norm{\psi_{1}}} \nonumber \\
  &=& \lambda_{\sz} \avg{A_{1}} \,,
\end{IEEEeqnarray}
where $\lambda_{\sz} \in [-1, 1]$ is related to the symmetrization-step
probability by $\lambda_{\sz} = \bar{p} - p$, and that
\begin{IEEEeqnarray}{rCl}
  \braket{\chi_{=}}{\chi_{\neq}}
  &=& \bar{p} \braket{\psi_{0}}{\psi'_{1}} \braket{\phi_{0}}{\phi_{1}}
  + p \braket{\psi'_{1}}{\psi_{0}} \braket{\phi_{1}}{\phi_{0}} \nonumber \\
  &=& \lambda_{\sx} \,
  e^{i \varphi} \bra{\psi_{0}} B \otimes \id_{E} \ket{\psi_{1}}
  \nonumber \\
  &=& \lambda_{\sx} \, \babs{
    \re \bigsq{\bra{\psi_{0}} B \otimes \id_{E} \ket{\psi_{1}}}} \,,
\end{IEEEeqnarray}
where we recall that we set $\braket{\phi_{0}}{\phi_{1}} = \lambda_{\sx} \in [0, 1]$, while
\begin{equation}
  \avg{\sx \otimes B}
  = 2 \re \bigsq{\bra{\phi_{0}} B \otimes \id_{E} \ket{\phi_{1}}} \,,
\end{equation}
so that
\begin{equation}
  2 \braket{\chi_{=}}{\chi_{\neq}}
  = \lambda_{\sx} \, \abs{\avg{\sx \otimes B}} \,.
\end{equation}
Putting all this together and recalling that we identify $A_{1}$ with $\sz$, and can choose $\bar{A}_{1} = \sx$, means that we finally get
\begin{equation}
  H(A_{1} | E)
  \geq g_{q} \bigro{\lambda_{\sz} \avg{A_{1}},\,
    \lambda_{\sx} \abs{\avg{\bar{A}_{1} \otimes B}}}
\end{equation}
for all $-1 \leq \lambda_{\sz}, \lambda_{\sx} \leq 1$ (as the derivation we
have given applies for any values of the symmetrization probability $p$ and
overlap $\braket{\phi_{0}}{\phi_{1}}$ we may wish to use). This confirms that
the inequality
\begin{equation}
  H(A_{1} | E) \geq g_{q}(Z, X)
\end{equation}
holds for any (real) numbers satisfying
\begin{IEEEeqnarray}{c+t+c}
  \abs{Z} \leq \abs{\avg{A_{1}}} &and&
  \abs{X} \leq \abs{\avg{\bar{A}_{1} \otimes B}} \,. \IEEEeqnarraynumspace
\end{IEEEeqnarray}

\section{Analytic solution for $p = 1/2$}
\label{ap:two-basis}

Here we derive in detail the average entropy bound for the two-basis protocol in the case that Alice's measurements are used equiprobably. When $p = 1/2$, the minimization problem~\eqref{eq:twobasis-polyopt} in Section~\ref{sec:correlations} simplifies to
\begin{IEEEeqnarray}{u+rCl}
  \label{eq:analytical-opt-equiprobable}
  minimize & \IEEEeqnarraymulticol{3}{l}{
    f(\lambda, \mu, \varphi_A)
    = \sin \bigro{\tfrac{\varphi_A}{2}}^2 \lambda^2
    + \cos \bigro{\tfrac{\varphi_A}{2}}^2 \mu^2} \nonumber \\
  subject to & \babs{\cos \bigro{\tfrac{\varphi_A}{2}}} \abs{\lambda}
  + \babs{\sin \bigro{\tfrac{\varphi_A}{2}}} \abs{\mu}
  &\geq& S/2 \nonumber \\
  & \lambda^2 &\leq& 1 \nonumber \\
  & \mu^2 &\leq& 1 \,,
\end{IEEEeqnarray}
where we have reintroduced the angle $\varphi_{A}$ from earlier in the section explicitly and used that the single constraint involving the variable $\Delta$ becomes irrelevant. As we stated in Section~\ref{sec:correlations} and show here, the above problem can be solved analytically subject to finding the root of a degree four polynomial.

In the following, we will assume that $S > 2$, since the solution to the classical case $S = 2$ is trivially $E\du{\frac{1}{2}}{2}=0$. 

First, we note that, as our problem is invariant under the transformations $\lambda \mapsto - \lambda$ and $\mu \mapsto -\mu$ and that, for $S > 2$, the points $\mu = 0$ or $\lambda = 0$ do not satisfy the first constraint \begin{equation}
  \label{eq:analytical-chsh-constraint}
  \babs{\cos \bigro{\tfrac{\varphi_A}{2}}} \abs{\lambda}
  + \babs{\sin \bigro{\tfrac{\varphi_A}{2}}} \abs{\mu}
  \geq S/2 \,,
\end{equation}
we can replace the constraints $\lambda^{2} \leq 1$ and $\mu^{2} \leq 1$ with $0 < \lambda \leq 1$ and $0 < \mu \leq 1$.

Moreover, the problem is also invariant under the transformation $\varphi_A \mapsto 2 \pi-\varphi_A$, meaning that for all solutions such that $\varphi_A\in [0,\pi]$, there exists an equivalent solution in $[\pi, 2\pi]$. Thus, we can restrict the domain of $\varphi_A$ to be $0 < \varphi_A < \pi$, where we excluded the boundaries since the cases $\varphi_A = 0, \pi$ are not in agreement with $S>2$.  

The function that we need to minimize can be rewritten as
\begin{IEEEeqnarray}{rCl}
  f(\lambda,\mu,\varphi_A) 
  &=& \frac{\lambda^2}{2} \bigro{1 - \cos(\varphi_A)}
  + \frac{\mu^2}{2} \bigro{1 + \cos(\varphi_A)} \,. \nonumber \\*
\end{IEEEeqnarray} 

Let us look for a minimum for our function by checking where its derivatives are zero. We start with
\begin{equation}
  \frac{d}{d\mu} f(\lambda, \mu, \varphi_A)
  = \mu \bigro{1 + \cos(\varphi_A)} \,.
\end{equation}
Here, $\frac{d}{d\mu} f(\lambda,\mu,\varphi_A) = 0$ if and only if $\mu = 0$ or $\varphi_A = \pi$. These points are not part of the restricted domain that we are considering. We conclude that the minimum must be at the boundaries of our domain. From now on, we will analyze this case. 

\paragraph*{Case 1:} We consider the boundary $\lambda = 1$. We have
\begin{equation}
  f(1,\mu,\varphi_A) = \frac{1 + \mu^2}{2}
  + \frac{\cos(\varphi_A)}{2} (\mu^2-1)
\end{equation}
and
\begin{equation}
  \frac{d}{d\mu } f(1,\mu,\varphi_A)
  = \mu \bigro{1 + \cos(\varphi_A)} \,,
\end{equation}
thus $\frac{d}{d\mu}f(\lambda, \mu, \varphi_A) = 0$ if and only if $\mu = 0$ or $\varphi_A = \pi$. Such solutions are not in the domain.

\paragraph*{Case 2:} We consider the boundary $\mu = 1$.
Analogously, we obtain non-feasible solutions.

\paragraph*{Case 3:} We consider the boundary $\cos \bigro{\tfrac{\varphi_A}{2}} \lambda + \sin \bigro{\frac{\varphi_A}{2}} \mu = S/2$. 
This region is the one in which
\begin{IEEEeqnarray}{rCl}
  \mu_{*} &=& \lambda \frac{\sin(\varphi_A)}{\cos(\varphi_A)-1}
  - S \frac{\sin \bigro{\tfrac{\varphi_A}{2}}}{\cos(\varphi_A)-1} \nonumber \\
  &=& \lambda \frac{\sqrt{1-x^2}}{x-1}
  - \frac{S}{\sqrt{2}} \frac{\sqrt{1-x}}{x-1} \,,
\end{IEEEeqnarray}
where we made the change of variable $x = \cos(\varphi_A)$. The domain of $x$ is $-1 < x < 1$. 

We have
\begin{equation}
  f(\lambda, \mu_{*}, x) = \lambda^2 \frac{1 + x^2}{1 - x}
  - \lambda \frac{S (1 + x)^{3/2}}{\sqrt{2} (1 - x)}
  + \frac{S^2 (1 + x)}{4 (1 - x)}
\end{equation}
and
\begin{equation}
  \frac{d}{d\lambda} f(\lambda, \mu_{*}, x)
  = 2 \lambda \frac{1 + x^2}{1 - x}
  - \frac{S (1 + x)^{3/2}}{\sqrt{2} (1-x)} \,.
\end{equation}
Now, recalling that we assumed $x \neq 1$, we have that $\frac{d}{d\lambda} f(\lambda, \mu_{*}, x) = 0$ iff
\begin{equation}
  \lambda = \frac{S (x+1)^{3/2}}{2\sqrt{2} (x^2 + 1)} = \lambda_{*} \,.
\end{equation}
Thus,
\begin{equation}
  f(\lambda_{*}, \mu_{*}, x) = \frac{S^2}{8} \frac{1 - x^2}{1 + x^2} \,,
\end{equation} which is a concave function of $x$, meaning the minimum is at the intersection between boundaries. 

\paragraph*{Case 3+1:} We intersect the boundary of case 3 with $\lambda = 1$.
We get 
\begin{equation}
  \mu_{*} = \frac{\sqrt{1-x^2}}{x-1}
  - \frac{S}{\sqrt{2}} \frac{\sqrt{1-x}}{x-1} \,.
\end{equation}
Here, requiring $\mu_{*} \leq 1$, we obtain the condition
\begin{equation}
  \label{eq:constr-analyt}
  - \frac{S}{4}\sqrt{8-S^2} \leq x \leq \frac{S}{4} \sqrt{8-S^2} \,.
\end{equation}
We have
\begin{equation}
  f(1, \mu_{*}, x) = \frac{x^2 + 1}{1 - x}
  - \frac{S (x + 1)^{3/2}}{\sqrt{2} (1 - x)}
  + \frac{S^2 (x + 1)}{4 (1 - x)}
\end{equation}
and 
\begin{IEEEeqnarray}{l}
  \frac{d}{dx} f(1, \mu_{*}, x) =  \nonumber \\
  \quad {} \frac{4x(2-x) + 2(S^2+2) + S (x-5) \sqrt{2(1+x)}}{4(x-1)^2} \,, 
  \IEEEeqnarraynumspace
\end{IEEEeqnarray}
hence, since $x \neq 1$, $\frac{d}{dx} f(1, \mu_{*}, x) = 0$ iff 
\begin{equation}
  \label{eq:analyt-to-solve}
  4x(2-x) + 2(S^2+2) + S(x-5) \sqrt{2(1+x)} = 0 \,.
\end{equation}

\paragraph*{Case 3+2:} We intersect the boundary of case 3 with $\mu = 1$. Here, one can check that we obtain the same result as in case 3+1.

\paragraph*{Case 1+2:} We consider $\lambda = \mu = 1$. With this choice we have $E\du{\frac{1}{2}}{2} = 1$ $\forall \varphi_A$. This region of parameters does not contain in general the absolute minimum.

We conclude that the solution to the optimization problem must be the one of case 3+1 (or equivalently 3+2). If there is more than one solution to Eq.~\eqref{eq:analyt-to-solve} satisfying the constraints~\eqref{eq:constr-analyt}, then we take the smallest one. 

We used Mathematica to find the roots of Eq.~\eqref{eq:analyt-to-solve} analytically. Moreover, imposing the constraints~\eqref{eq:constr-analyt} and $S > 2$, we found a single solution. We used the resulting expression for the computations for $p = 1/2$ done in Section~\ref{sec:two-basis}.

\section{Optimality of CHSH for the two-basis protocol}
\label{ap:twobasis-chsh}

In the case that the bases are used equiprobably, i.e., $p = 1/2$, the symmetries of the two-basis DIQKD protocol studied in section~\ref{sec:two-basis} imply that the CHSH Bell expectation value alone already gives the optimal bound on the average conditional entropy
\begin{equation}
  H(A_{x} | X E)
  \propto \tfrac{1}{2} H(A_{1} | E) + \tfrac{1}{2} H(A_{2} | E)
\end{equation}
for the optimal CHSH-violating correlations attenuated by white noise. The reason for this is that, given any quantum strategy giving a particular value of the average entropy and CHSH expectation value, one can construct a new symmetrized strategy giving the same entropy and CHSH expectation value.

To see this, let us suppose we have a particular quantum strategy $\mathcal{Q} = (\rho_{ABE}, A_1, A_2, B_1, B_2)$. We note first that both conditional entropies $H(A_{x} | E)$ and the CHSH expectation value $S = \avg{A_{1} B_{1}} + \avg{A_{1} B_{2}} + \avg{A_{2} B_{1}} - \avg{A_{2} B_{2}}$ are unchanged if we flip all the measurements, i.e., do $A_{x} \mapsto -A_{x}$ and $B_{y} \mapsto -B_{y}$. By randomly and equiprobably using these two strategies we can force Alice's and Bob's local outcomes to become equiprobable. This corresponds to using a new strategy $\mathcal{Q}' = (\rho'_{ABE}, A'_1, A'_2, B'_1, B'_2)$ with
\begin{IEEEeqnarray}{rCl}
  A'_{x} &=& A_{x} \oplus -A_{x} \,, \\
  B'_{y} &=& B_{y} \oplus -B_{y} \,, \\
  \rho'_{ABE} &=& \tfrac{1}{2} \rho_{ABE} \oplus \tfrac{1}{2} \rho_{ABE} \,,
\end{IEEEeqnarray}
for which the CHSH expectation value and the values of the entropies are unchanged, but for which $\avg{A'_{x}} = \avg{B'_{y}} = 0$.

Next, we use that the average entropy and CHSH both remain unchanged under the two transformations
\begin{IEEEeqnarray}{c+c}
  T_{1} : \left\{ \begin{IEEEeqnarraybox}[][c]{rCl}
    A_{1} &\mapsto& A_{1} \\
    A_{2} &\mapsto& -A_{2} \\
    B_{1} &\mapsto& B_{2} \\
    B_{2} &\mapsto& B_{1}
  \end{IEEEeqnarraybox} \right. &
  T_{2} : \left\{ \begin{IEEEeqnarraybox}[][c]{rCl}
    A_{1} &\mapsto& A_{2} \\
    A_{2} &\mapsto& A_{1} \\
    B_{1} &\mapsto& B_{1} \\
    B_{2} &\mapsto& -B_{2}
  \end{IEEEeqnarraybox} \right. \,,
\end{IEEEeqnarray}
as well as their composition $T_{2} \circ T_{1}$. By randomly using the strategy $\mathcal{Q}'$ with neither, either one, or both transformations applied, we construct a new strategy $\mathcal{Q}'' = (\rho''_{ABE}, A''_1, A''_2, B''_1, B''_2)$ with
\begin{IEEEeqnarray}{rCl}
  A''_{1} &=& A'_{1} \oplus A'_{1} \oplus A'_{2} \oplus A'_{2} \,, \\
  A''_{2} &=& A'_{2} \oplus -A'_{2} \oplus A'_{1} \oplus -A'_{1} \,, \\
  B''_{1} &=& B'_{1} \oplus B'_{2} \oplus B'_{1} \oplus -B'_{2} \,, \\
  B''_{2} &=& B'_{2} \oplus B'_{1} \oplus -B'_{2} \oplus B'_{1} \,, \\
  \rho''_{ABE} &=& \tfrac{1}{4} \rho'_{ABE} \oplus \tfrac{1}{4} \rho'_{ABE}
  \oplus \tfrac{1}{4} \rho'_{ABE} \oplus \tfrac{1}{4} \rho'_{ABE} \,,
  \IEEEeqnarraynumspace
\end{IEEEeqnarray}
for which
\begin{equation}
  \avg{A''_{1} B''_{1}} = \avg{A''_{1} B''_{2}}
  = \avg{A''_{2} B''_{1}} = - \avg{A''_{2} B''_{2}} = S/4 \,.
\end{equation}

As, given any strategy $\mathcal{Q}$, we can in this way always construct a strategy $\mathcal{Q}''$ with the same average entropy and CHSH expectation value, but satisfying $\avg{A''_{x}} = \avg{B''_{y}} = 0$ and $\avg{A''_{1} B''_{1}} = \avg{A''_{1} B''_{2}} = \avg{A''_{2} B''_{1}} = -\avg{A''_{2} B''_{2}}$, we can infer that these constraints, if they are satisfied for real correlations, do not contain any information other than the CHSH expectation value that can be used to improve the entropy bound.

\section{Explicit attack for the two-basis protocol}
\label{sec:attack-two-basis}

We describe here an explicit attack for the two-basis protocol in the case $p = 1/2$, which we conjecture to be optimal.

Suppose that Alice, Bob, and Eve share the optimal symmetric BB84 attack state
\begin{IEEEeqnarray}{rll}
  \ket{\Psi}_{ABE} = \frac{1}{2} \Bigsq{&
    (1 + E) \ket{\phi^{+}}_{AB} \ket{++}_{E} & \nonumber \\
    &+\> \sqrt{1 - E^{2}} \ket{\phi^{-}}_{AB} \ket{+-}_{E} & \nonumber \\
    &+\> \sqrt{1 - E^{2}} \ket{\psi^{+}}_{AB} \ket{-+}_{E} & \nonumber \\
    &+\> (1 - E) \ket{\psi^{-}}_{AB} \ket{--}_{E} &} \,,
\end{IEEEeqnarray}
where $\ket{\phi^{\pm}}$ and $\ket{\psi^{\pm}}$ are the four Bell states, depending on some number $0 \leq E \leq 1$. Its marginal once Eve is traced out is
\begin{equation}
  \Psi_{AB} = \frac{1}{4} \Bigsq{\id \otimes \id + E \, \sx \otimes \sx
    - E^{2} \, \sy \otimes \sy + E \, \sz \otimes \sz} \,.
\end{equation}
By measuring $A_{1} = \sz$, $A_{2} = \sx$, and $B_{1,2} = (\sz \pm \sx)/\sqrt{2}$, the highest possible CHSH expectation value of $S = 2 \sqrt{2} E$ with this state is obtained. Direct computation of the conditional entropies after Alice measures $\sz$ and $\sx$ on this state gives
\begin{equation}
  \label{eq:twobasis-conjecture-highS}
  \tfrac{1}{2} H(A^q_{1} | E) + \tfrac{1}{2} H(A^q  _{2} | E)
  = f_{q} \bigro{S / \sqrt{8}}
\end{equation}
where $f_{q}$ is the same BB84 bound with noisy preprocessing used earlier and given by Eq.~\eqref{eq:fqx}. This is too high to be the optimal bound on the average entropy for all $S$, as the correct bound must attain $h(q)$ at $S = 2$. But we can construct a plausible strategy by taking a convex mixture (similar to the construction in Section~2 of \cite{ref:wap2021}) of the strategy just described with a deterministic one giving $\bigro{H(A^q_X|XE),\, S} = (h(q),\, 2)$. This gives
\begin{equation}
  \tfrac{1}{2} H(A^q_{1} | E) + \tfrac{1}{2} H(A^q_{2} | E)
  = \bar{f}_{q} \bigro{S / \sqrt{8}} \,,
\end{equation}
where
\begin{equation}
  \bar{f}_{q}(x) = \begin{cases}
    f_{q}(x) &\text{if } x \geq x_{*} \\
    h(q) + f'_{q}(x_{*}) (x - 1/\sqrt{2}) &\text{if } x \leq x_{*}
  \end{cases}
\end{equation}
with $x_{*}$ (dependent on $q$) such that
\begin{equation}
  h(q) + f'_{q}(x_{*}) (x - 1/\sqrt{2}) = f_{q}(x_{*}) \,.
\end{equation}

\section{Explicit attack saturating the qubit entropy bound with bias \eqref{eq:bias}}
\label{sec:attack-bias}

One can verify that the qubit bound \eqref{eq:bias} is attained with measurements and an initial state of the form
\begin{IEEEeqnarray}{rCl}
  A_{1} &=& \sz \,, \\
  A_{2} &=& \sx \,, \\
  B_{1} &=& \cos \bigro{\tfrac{\varphi_{B}}{2}} \sz
  + \sin \bigro{\tfrac{\varphi_{B}}{2}} \sz \,, \\
  B_{2} &=& \cos \bigro{\tfrac{\varphi_{B}}{2}} \sz
  - \sin \bigro{\tfrac{\varphi_{B}}{2}} \sz \,,
\end{IEEEeqnarray}
and
\begin{equation}
  \ket{\Psi}_{ABE}
  = \cos \bigro{\tfrac{\theta}{2}} \ket{00}_{AB} \ket{\psi_{0}}_{E}
  + \sin \bigro{\tfrac{\theta}{2}} \ket{11}_{AB} \ket{\psi_{1}}_{E}
\end{equation}
with
\begin{IEEEeqnarray}{rCl}
  \label{eq:bias-opt-costheta}
  \cos(\theta) &=& \avg{A_{1}} \,, \\
  \label{eq:bias-opt-sinthetaF}
  \sin(\theta) \braket{\psi_{0}}{\psi_{1}} &=& \sqrt{S^{2}/4 - 1} \,, \\
  \cos \bigro{\tfrac{\varphi_{B}}{2}} &=& 2/S \,, \\
  \sin \bigro{\tfrac{\varphi_{B}}{2}} &=& \sqrt{1 - 4/S^{2}} \,.
\end{IEEEeqnarray}
Note that, because $\cos(\theta)^{2} + \sin(\theta)^{2} \abs{\braket{\psi_{0}}{\psi_{1}}}^{2} \leq 1$, \eqref{eq:bias-opt-costheta} and \eqref{eq:bias-opt-sinthetaF} are only consistent with each other if
\begin{equation}
  \avg{A_{1}}^{2} + S^{2}/4 \leq 2 \,,
\end{equation}
but this is a known boundary of the quantum set \cite{ref:pam2010,ref:amp2012}.

\end{document}

%% file: preamble.tex
\usepackage{graphicx}
\graphicspath{{figs/}}
\usepackage{tikz}
\usepackage{amsmath}
\usepackage{amssymb}
\usepackage{amsthm}
\usepackage{bbm}
\usepackage{float}
\usepackage{IEEEtrantools}
\usepackage[bb=boondox]{mathalfa}
\usepackage{microtype}
\usepackage{quant_defs}

\definecolor{mygrey}{gray}{0.35}
\definecolor{myblue}{rgb}{0.2,0.2,0.8}
\definecolor{myzard}{cmyk}{0,0,0.05,0}
\definecolor{mywhite}{rgb}{1,1,1}
\definecolor{myred}{rgb}{0.9,0.1,0.}
\usepackage[colorlinks=true,citecolor=myblue,linkcolor=myblue,urlcolor=myblue]{hyperref}

\DeclareMathOperator{\re}{Re}
\DeclareMathOperator{\verx}{Vert}

\newcommand{\vect}[1]{\boldsymbol{#1}}

\newcommand{\sx}{\mathrm{X}}
\newcommand{\sy}{\mathrm{Y}}
\newcommand{\sz}{\mathrm{Z}}

\newcommand{\rxx}{\mathrm{xx}}
\newcommand{\rxz}{\mathrm{xz}}
\newcommand{\rzx}{\mathrm{zx}}
\newcommand{\rzz}{\mathrm{zz}}

\newcommand{\hilb}[1]{\mathcal{H}_{#1}}

\makeatletter
\newcommand{\pushright}[1]{\ifmeasuring@#1\else\omit\hfill$\displaystyle#1$\fi\ignorespaces}
\makeatother

\theoremstyle{plain} % just in case the style had changed
\newcommand{\thistheoremname}{}
\newtheorem*{genericthm}{\thistheoremname}
\newenvironment{namedthm}[1]
  {\renewcommand{\thistheoremname}{#1}%
   \begin{genericthm}}
  {\end{genericthm}}

\input{journals}
\usepackage[numbers,sort&compress]{natbib}